\PassOptionsToPackage{svgnames,table,xcdraw}{xcolor}

\documentclass[preprint,journal]{vgtc}            

\onlineid{0}

\vgtccategory{Research}

\vgtcpapertype{application/design study}

\title{Data Type Agnostic Visual Sensitivity Analysis}

\author{%
  \authororcid{Nikolaus Piccolotto}{0000-0001-6876-6502},
  \authororcid{Markus Bögl}{0000-0002-8337-4774},
  \authororcid{Christoph Muehlmann}{0000-0001-7330-8434},
  \authororcid{Klaus Nordhausen}{0000-0002-3758-8501},\\
  \authororcid{Peter Filzmoser}{0000-0002-8014-4682},
  \authororcid{Johanna Schmidt}{0000-0002-9638-6344},
 and \authororcid{Silvia Miksch}{0000-0003-4427-5703}
}

\authorfooter{
  \item
    Nikolaus Piccolotto, Markus Bögl, Christoph Muehlmann, Peter Filzmoser and Silvia Miksch are with TU~Wien, Austria.
  	E-mail: \{firstname\}.\{lastname\}@tuwien.ac.at.

  \item Klaus Nordhausen is with University of Jyväskylä, Finland.
  	E-mail: klaus.k.nordhausen@jyu.fi.

   \item
  	Johanna Schmidt is with VRVis GmbH, Austria.
  	E-mail: johanna.schmidt@vrvis.at.
}

\abstract{%
Modern science and industry rely on computational models for simulation, prediction, and data analysis. Spatial blind source separation (SBSS) is a model used to analyze spatial data. Designed explicitly for spatial data analysis, it is superior to popular non-spatial methods, like PCA. However, a challenge to its practical use is setting two complex tuning parameters, which requires parameter space analysis. In this paper, we focus on sensitivity analysis (SA). SBSS parameters and outputs are spatial data, which makes SA difficult as few SA approaches in the literature assume such complex data on both sides of the model. Based on the requirements in our design study with statistics experts, we developed a visual analytics prototype for data type agnostic visual sensitivity analysis that fits SBSS and other contexts. The main advantage of our approach is that it requires only dissimilarity measures for parameter settings and outputs (\Fref{fig:teaser}). We evaluated the prototype heuristically with visualization experts and through interviews with two SBSS experts. In addition, we show the transferability of our approach by applying it to microclimate simulations. Study participants could confirm suspected and known parameter-output relations, find surprising associations, and identify parameter subspaces to examine in the future. During our design study and evaluation, we identified challenging future research opportunities.

}

\keywords{Visual analytics, parameter space analysis, sensitivity analysis, spatial blind source separation.}

\teaser{
  \centering
  \includegraphics[width=\linewidth]{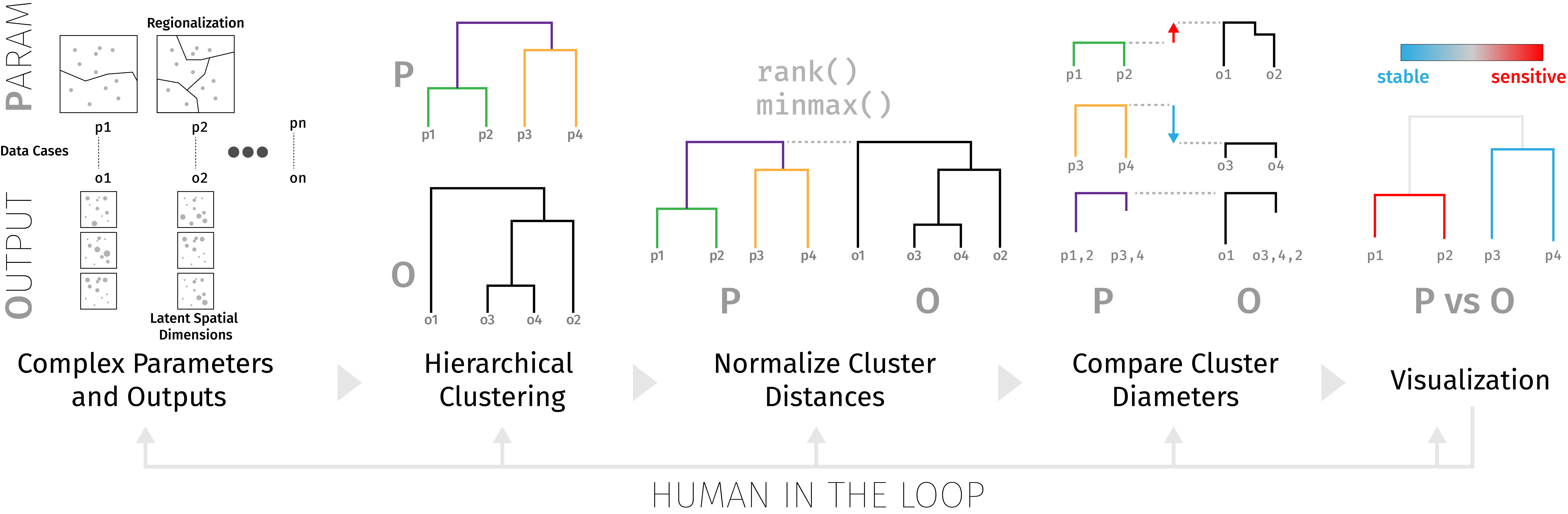}
  \caption{%
  	Illustration of the visualization pipeline used for the \dde. The key idea of it is to use cluster diameters as a measure for variation in complex parameter settings and outputs. Given dissimilarity measures for each, we perform hierarchical clustering separately in each space. Distances are normalized by ranking or min-max normalization for comparison. For every cluster obtained through hierarchical clustering, we evaluate the difference in its diameter and visualize that by color.  %
  }
  \label{fig:teaser}
}

\graphicspath{{figs/}{figures/}{pictures/}{images/}{./}} 

\usepackage{mathptmx}                  
\usepackage{csquotes}
\usepackage[plain]{fancyref}
\usepackage{bm}
\usepackage{MnSymbol}
\usepackage{amsfonts}
\usepackage{booktabs}
\usepackage{textcomp}
\usepackage[labelfont=bf,textfont=it]{caption}
\usepackage{subcaption}
\usepackage{tikz}
\usepackage{pgfplots}
\usepackage[linesnumbered]{algorithm2e}
\pgfplotsset{width=.5\linewidth}

\newcommand{\jv}{SE2 }
\newcommand{\jve}{SE2}
\newcommand{\cc}{SE1 }
\newcommand{\cce}{SE1}
\newcommand{\dd}{Discrepancy Dendrogram }
\newcommand{\dde}{Discrepancy Dendrogram}
\newcommand{\js}{JS }
\newcommand{\jse}{JS}
\newcommand{\mv}{ME1 }
\newcommand{\mve}{ME1}
\newcommand{\np}{NP }
\newcommand{\npe}{NP}

\newcommand{\sae}{S\kern-0.025em A}
\newcommand{\sa}{S\kern-0.025em A }
\newcommand{\saie}{S\kern+0.025em A}
\newcommand{\sai}{S\kern+0.025em A }

\usetikzlibrary{calc}
\usetikzlibrary{%
  arrows,%
  shapes,
  chains,%
  matrix,%
  positioning,
  scopes,%
  decorations.pathmorphing,
  shadows,%
  calc
}\tikzset{
     textWcolor/.style={
        		circle,
            fill=white, 
            draw=black, 
            line width=1pt,
            font=\sffamily\scriptsize,
            opacity=0.7,
            inner sep=0pt,
            minimum size=10pt
    },
    textWborder/.style={
        		circle,
            fill=none,
            draw=black, 
            line width=1pt,
            font=\normalfont\bfseries\small,
            opacity=0.7,
            inner sep=0pt,
            minimum size=12pt
    },
    textOnly/.style={
        		rectangle,
        		fill=none, 
            font=\normalfont\scriptsize,
            inner sep=0pt,
            minimum size=10pt
    },
    cycleLabels/.style={
        		rectangle,
            fill=none,
            draw=none,
            font=\normalfont\small,
            inner sep=0pt,
            anchor=west
    },
    mathnode/.style={
        		rectangle,
            fill=none,
            draw=none,
            font=\normalfont\normalsize,
            inner sep=0pt,
            anchor=west
    },
    sfnode/.style={
        		rectangle,
            fill=none,
            draw=none,
            font=\sffamily\small,
            inner sep=0pt,
            anchor=west
    },
    image overlay index/.style={
        		circle,
            fill=col2,
            draw=black!40,
            line width=0.5pt,
            font=\normalfont\scriptsize,
            opacity=0.7,
            inner sep=0pt,
            minimum size=10pt
    }
}

\newcommand{\glyphSenRede}[0]{\includegraphics[height=2mm]{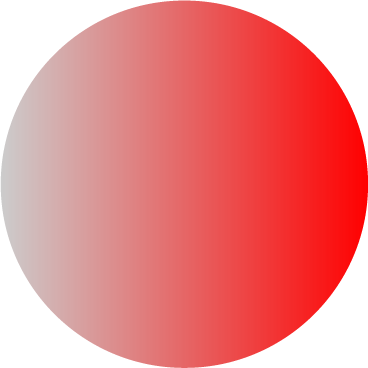}}

\newcommand{\glyphSenBluee}[0]{\includegraphics[height=2mm]{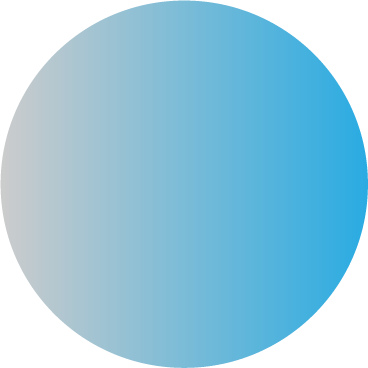}}

\newcommand{\ddshorte}[2]{\underline{#1}#2}

\begin{document}


\firstsection{Introduction}
\label{sec:introduction}

\maketitle

In many domains, data analysis requires dealing with multivariate measurements in space. For instance, mining corporations and public agencies may analyze geochemical soil samples for mine prospecting or investigating environmental pollution, respectively. Depending on the specific goal and application, various tasks, e.g., dimension reduction or finding meaningful linear combinations of variables, must be carried out on such datasets. Spatial blind source separation (SBSS) \cite{nordhausen2015, bachoc2020, muehlmann2022} is designed explicitly for multivariate spatial data and reveals linear combinations of such data. SBSS offers various benefits compared to alternative methods, e.g., it keeps the well-known loadings-scores scheme from principal component analysis and adequately accounts for spatial dependence due to its model-based approach. Therefore, latent dimensions identified with SBSS often correspond to the physical reality where data was collected, making it an excellent analysis tool for spatial data. A detailed description of SBSS is out of scope for this paper, and we refer interested readers to \cite{nordhausen2015, piccolotto2022, muehlmann2022}. SBSS has been successfully applied to a geochemical dataset \cite{nordhausen2015} and may be potentially used in any application domain that involves multivariate quantitative measurements at different locations.

\begin{figure}[tb]
    \centering
    \includegraphics[width=.75\linewidth]{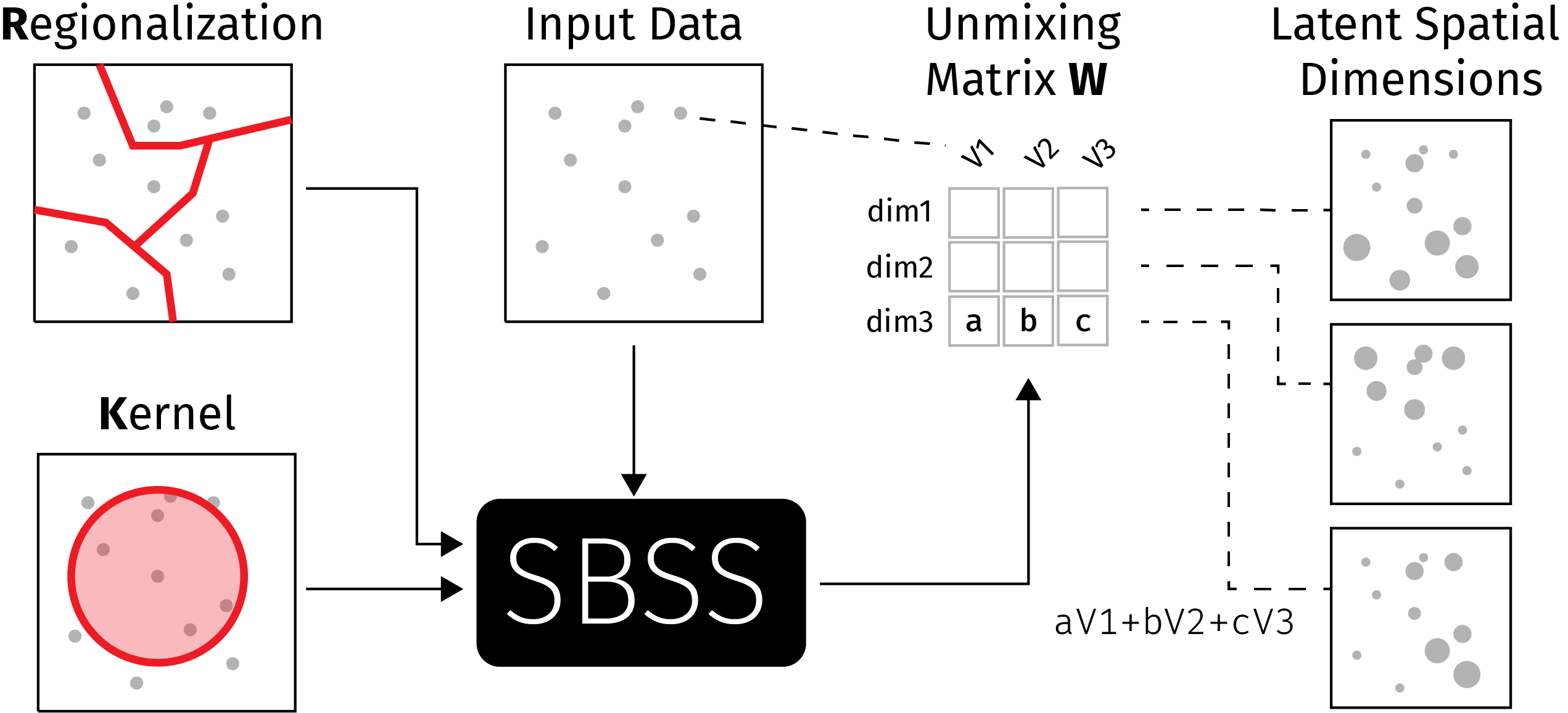}
    \caption{SBSS \cite{nordhausen2015, piccolotto2022, muehlmann2022} takes a regionalization (R) and a kernel (K) as parameters and outputs a linear combination of input variables (latent spatial dimensions), described by the unmixing matrix (W).}
    \label{fig:sbss-model}
\end{figure}

SBSS requires setting two complex tuning parameters: A partition of the spatial domain in non-overlapping regions (regionalization) and a ring-shaped point neighborhood (kernel). On the other side of the model (\Fref{fig:sbss-model}), SBSS yields a set of latent spatial dimensions (i.e., maps), where each is a linear combination of original dimensions with weights (loadings) given by the unmixing matrix. Consequently, parameter space analysis tasks \cite{sedlmair2014} become relevant. Previous work \cite{piccolotto2022} focused on the \emph{optimization} task, but \emph{sensitivity analysis} (\sae) is considered equally important for SBSS. \sa compares the relative variation in parameter settings and output of the model, thus highlighting relevant/irrelevant parameters and their stable/sensitive ranges. This analysis is essential to obtain and communicate reliable results, i.e., those not a consequence of luck and coincidence. \sa is especially important for SBSS as it lacks so far any goodness-of-fit criteria; hence deciding between alternative parameter settings is challenging. \sa can help with this decision as in prior work on blind source separation \cite{piccolotto2022a, piccolotto2022}, analysts noted that they find stable parameter settings more trustworthy and associated outputs more likely to be the \enquote{real} solution. \sa may thus further strengthen the outcome of an \emph{optimization} task and, additionally, inform geostatistical modeling: If, e.g., the regionalization parameter barely influences the output, analysts might reasonably suspect that the input dataset is spatially stationary (a geostatistical modeling decision).

SBSS is interesting for the visualization community primarily because of the mentioned affordances of its parameters and outputs: Parameter settings and outputs are spatial objects or otherwise complex in a way that a multivariate representation does not do them justice. While the literature contains many examples of visual parameter space exploration \cite{sedlmair2014,piccolotto2023}, to the best of our knowledge, none of them support complex parameters and outputs without resorting to multivariate representation or feature derivation (\Fref{sec:related-work}). However, these requirements are not specific to SBSS, as many examples exist for models with complex parameters \emph{and} outputs. For instance, spatial or time-varying inputs and outputs can arise in microclimate simulations \cite{vuckovic2022a}. They predict meteorological variables (e.g., air temperature, humidity, or wind speed) in a small area, typically for a single street or building.

We intend to close this gap with our paper. The core idea of our proposal is illustrated in \Fref{fig:teaser}: We take a cluster's diameter as a measure of variation for the contained parameter settings or associated outputs (referred to as data cases, respectively). Then we can enable \sa for SBSS in the following way. Given appropriate dissimilarity measures for data cases, we compute pairwise distances in each space (parameter and output), based on which a hierarchical clustering is produced. After normalizing distances, we compute the diameter difference of all clusters between one space and the another. This information is then presented in our main visualization, the \dde. Supporting visualizations complete required user tasks. In particular, the contributions of our design study are that we

\begin{itemize}
    \item propose a task abstraction for \sa in the context of SBSS (\Fref{sec:task-abstraction});
    \item based on SBSS requirements, develop a visualization that supports \sa and works on any data type (\Fref{fig:teaser}, \Fref{sec:discrepancy-dendrogram});
    \item integrate this and other visualizations in a visual analytics prototype (\Fref{sec:visualizations});
    \item evaluate the prototype with experts in visualization (\Fref{sec:evaluation:vis}) and SBSS (\Fref{sec:evaluation:sbss});
    \item show the transferability to other problems by applying our approach to microclimate simulations (\Fref{sec:evaluation:vrvis}).
\end{itemize}

\section{Related Work}
\label{sec:related-work}

\subsection{Sensitivity Analysis}
\label{sec:related-work:sa}

Sensitivity Analysis (\sae) is \enquote{the study of how the uncertainty in the output of a model (numerical or otherwise) can be apportioned to different sources of uncertainty in the model input \cite[p.~1]{saltelli2002}.} \sa allows analysts to determine how variations in the input influence the output. A broad distinction between various \sa methods can be drawn at whether they are \emph{local} or \emph{global} \cite{saltelli2007}. Local methods are applicable when the model is linear as they yield, e.g., a partial derivative according to one parameter. An example of such local methods is the one-at-a-time approach, where one parameter is varied while the others are kept fixed. Global methods, on the other hand, are applicable to non-linear models, too. A well-known example is the Sobol index \cite{sobol1990}, a variance-based global \sa method. Several surveys exist \cite{hamby1994, ionescu-bujor2004, cacuci2004, iooss2015, borgonovo2016, saltelli2019} that collect and discuss both local and global methods. Methods covered in these surveys mainly consider models with multivariate parameters, e.g., the output scalar $y$ is a function of an input vector $\mathbf x$: $y = f(\mathbf x)$. Spatially-varying parameters \cite{lilburne2009, raimbault2018} or outputs \cite{marrel2011, ligmann-zielinska2013} have been considered as well. However, these methods do not fit to SBSS (\Fref{fig:sbss-model}).

\subsection{Visual Parameter Analysis}
\label{sec:related-work:vpa}

Visual parameter analysis (VPA) has a long history in the visualization literature, with seminal works published in the 1990s, like Design Galleries \cite{marks1997} or spreadsheet interfaces \cite{jankun-kelly2000}. Sedlmair et al. \cite{sedlmair2014} provided a common data flow model and a task taxonomy, such as optimization, uncertainty, or \sae. Piccolotto et al. \cite{piccolotto2023} surveyed user interfaces and visualizations that support visual parameter space exploration.
Several examples of VPA for multivariate parameters can be found in the literature \cite{yang2021a,knittel2021,cibulski2020,pajer2017,bergner2013,guo2011}. However, these approaches do not apply to SBSS parameters.
Many approaches have been used when it comes to visualizing parameter-output relations \cite{piccolotto2023}. When parameters are multivariate, visualizations that show correlations and trends can be used to carry out \sae, such as histograms, scatterplots, or PCPs \cite{beham2014, wang2017, cibulski2022}. These visualizations are often \emph{juxtaposed} and linked, such that selections in one view highlight the same data in other views \cite{matkovic2017}. Another option is to \emph{embed} parameters and outputs in the same visualization, e.g., by encoding them as axes in the same PCP \cite{steed2013} or by color-coding a 3D model \cite{doraiswamy2015}. A consequence of juxtaposition is that general visualization-independent approaches may be used together. E.g., first grouping data cases by similarity, then inspecting properties of individual groups \cite{bruckner2010, abuzuraiq2020, hazarika2020} is popular. Orban et al. \cite{orban2019} devised two linked dimensionally-reduced (DR) scatterplots, an approach that can generally be extended to complex data and SBSS parameters/outputs. However, our target users struggled with DR scatterplots in previous work \cite{piccolotto2022a}. The difficulty was that the DR spatializations looked like scatterplots but did not show the same information and required a different way of reading, which was unintuitive to them. Therefore, we developed an alternative approach. A more specific form of juxtaposition is to \emph{align} data cases in useful ways that highlight dependencies between parameters and outputs, e.g., as part of a spreadsheet \cite{luboschik2014,luboschik2015,eichner2020}. The idea is that dependencies become visible when the spreadsheet is sorted by multiple columns. However, it requires a compact visual representation. \emph{Superposition} may be possible if parameter and output refer to the same space, such as particle trajectories and their initial position \cite{gunther2016a}. \emph{Sequential Superposition} leverages a system's interactivity. The analyst may rapidly browse between parameter/output pairs, and sudden visual jumps in the emerging animation point to sensitive parameter ranges \cite{schulz2017, rojo2018, he2020}. Parameter and output visualizations may also be \emph{integrated} with explicit links drawn between them. E.g., a trapezoid that connects parameter and output histograms shows sensitivity by the relative length of horizontal segments \cite{weissenbock2016}. Another option for composite visualizations of parameters and outputs for \sa is \emph{nesting}, i.e., putting visualizations inside the marks of another, like correlation matrices in an interval tree \cite{eichner2020}.

Data mining methods may also support visual \sae. E.g., if regression analysis between parameter and output is possible, that information can be shown in the parameter visualization in the spirit of scented widgets \cite{willett2007, koyama2014, desai2019}. Correlation analysis between parameters and derived output features may also be done if they lend themselves to it \cite{eichner2020}. Developing a surrogate model augmenting the original model with fast but inaccurate output predictions for new parameter settings is standard practice in VPA \cite{sedlmair2014}. It may be possible to extract information from the surrogate to support \sae, such as parameters in linear regression \cite{matkovic2017}, or partial derivatives in neural networks \cite{hazarika2020}.

Generally, in existing work, either the parameter (by multivariate representation) or the output (by feature derivation) must have multivariate characteristics. Our contribution to visual sensitivity analysis enables it in situations where both parameter and output are of complex data types, e.g., spatial objects.

\subsection{Visual Cluster Analysis and Clustering Comparison}
\label{sec:related-work:clusters}


Clustering is an essential wide-spread class of data analysis methods, and various flavors were proposed over time \cite{xu2015}. Generally, clusterings partition data cases into coherent groups according to a distance function. Visual inspection of these groups may reveal previously hidden patterns. To visualize the whole clustering, nowadays, color-coded dimensionally-reduced scatterplots are commonly employed \cite{kwon2018, cavallo2019, xia2022}. However, these scatterplots are only approximate, as they contain projection errors \cite{jeon2022}, and may require specialized knowledge to interpret \cite{wattenberg2016}. Glyph-based visualizations \cite{cao2011} were proposed in the context of geospatial data. Dendrograms \cite{seo2002, galili2015} commonly depict hierarchical clusterings. Blanch et al. \cite{blanch2015} proposed the Dendrogramix, a combination of dendrogram and matrix visualization. The clustering outcome depends on the specific algorithm and parameters, so visualizations were proposed to compare these. However, they focus on the analysis of cluster members \cite{cavallo2019}, comparison of clusterings concerning parameters \cite{cavallo2019} or algorithms \cite{lyi2015, kwon2018}. I.e., the definition of distance between data cases is fixed. Our work may be seen as comparing clusterings with alternative distances (\Fref{fig:teaser}).

\section{Users \& Task Abstraction}
\label{sec:task-abstraction}

As in previous work on SBSS \cite{piccolotto2022}, our primary users are experts in statistics. We anticipate our user base to eventually include domain experts, e.g., from geochemistry. We conducted an extensive literature review \cite{piccolotto2023} to understand how visual VPA and, consequently, visual SA work in other contexts. Based on that, we distilled generic \sa sub-tasks to enable \sa on the SBSS-specific complex parameters with our clustering-based approach (T1--T5). We presented and discussed them with our collaborators (statistics/SBSS experts who are co-authors of this paper) to ensure their suitability. Based on these tasks, we developed the main visualization (\Fref{sec:discrepancy-dendrogram}).



\paragraph*{Tasks.} First, to start the analysis, analysts must \emph{compare the association between parameters and outputs} (T1). Pairs of highly associated parameters and outputs are less interesting to investigate. For any given parameter/output, they must \emph{assess its overall variation} (T2) to learn about contained similarity structures and outliers. Furthermore, analysts must \emph{identify groups of data cases with low/high variation in a parameter/output} (T3) in order to \emph{compare variation between parameters and outputs, both overall and for a group of data cases} (T4). To support analysts in reasoning why this variation happens, they must be able to \emph{view individual data cases } (T5).

\paragraph*{Guidelines.} In addition to user tasks, we formulate three design guidelines for the visualizations. These were informed by evaluations conducted in our past work \cite{piccolotto2022a,piccolotto2022} and by widely used visualization guidelines. First, \emph{visual marks of similar values should be adjacently arranged} (D1). This visual requirement suggests continuity that scalars exhibit naturally, but complex objects do not. It will make it easier to perceive stable/sensitive parameter ranges. \emph{Occlusion must be avoided} (D2) to not clutter the display. The visualization should, if possible, \emph{resemble a familiar graphic} (D3) that our target users are familiar with.


\section{\dde}
\label{sec:discrepancy-dendrogram}

We describe in this section how our main visualization, the \dde, is constructed (also compare \Fref{fig:teaser}). The complete VA prototype will be discussed in the following section. We aim for a visual-interactive approach for two reasons. First, we did not find numerical SA approaches that are applicable to our data (\Fref{sec:related-work:sa}). Second, our approach needs configuration (e.g., \Fref{sec:discrepancy-dendrogram:clustering} or \Fref{sec:discrepancy-dendrogram:distance-normalization}), where each choice highlights different patterns (compare \Fref{fig:evaluation:sbss:co-kw}), impacting the conclusions to draw. Thus, in an interactive setting, the analyst can quickly change between those configurations and thoroughly compare them (see, e.g., \Fref{sec:evaluation:sbss}).

The core of \sa is to compare the relative variation in parameter settings and outputs. It can readily be quantified for numbers (cf. variance-based \sa approaches), but measuring \emph{variation} for complex objects, like the spatial SBSS parameters, is not straightforward. Our proposal's core idea (\Fref{fig:teaser}) is to consider cluster diameters for that purpose: A cluster gets wider the more dissimilar contained data cases are. Conversely, the cluster diameter is zero when all contained data cases are the same. There are advantages to that approach. First of all, a clustering can be obtained when only pairwise similarity information (\Fref{sec:discrepancy-dendrogram:dissimilarity-measures}) is available. Thus a formal notion of variation need not exist for the data type at hand. Second, cluster analysis generally supports tasks T2 and T3 when one investigates global cluster structures (e.g., how many exist, how many data cases they contain) and local structures (e.g., finding outlier cases). Hence we propose to augment a visualization of cluster structures with the information required for \sae, i.e., whether clusters shrink or expand when applying another dissimilarity measure to the data cases. This approach can be seen as \emph{orienting} guidance \cite{ceneda2017} that points analysts to interesting data cases. The major available choices at this point are i) the type of visualization and ii) how to compute the augmenting information. The two choices are independent, and we focus on the latter before discussing the former in \Fref{sec:discrepancy-dendrogram:visualization}.

\paragraph*{Sampling.} Any parameter space analysis task requires a reasonable set of (parameter setting, output) tuples. Common desired sampling properties are that it is uniform and spans a large part of the parameter space, which is achieved via automated sampling techniques. These are hard problems for SBSS, where two random parameter settings are not a-priori equally reasonable. Domain knowledge critically informs parameter selection in SBSS \cite{piccolotto2022}. Single-execution runtimes measured in minutes or hours further complicate the issue. Thus, following study participants' current practices in SBSS and microclimate simulations, we rely on a few dozen, mostly manually selected, parameter settings and limit \sa insights to that subspace. While not solving everything at once, our approach still improves their current situation. 


\subsection{Dissimilarity Measures}
\label{sec:discrepancy-dendrogram:dissimilarity-measures}

Dissimilarity measures, considerably the basic requirement for any analysis, exist for many data types. A dissimilarity measure is a function $d(\cdot,\cdot) \rightarrow \mathbb{R}^+$ that quantifies how similar two objects are. Generally, we expect that $d(a,b) = 0$ iff $a=b$ and that $d(a,b)$ is strictly monotonically increasing with the differences between $a$ and $b$. We assume such a dissimilarity measure for every model parameter and output.


\subsection{Hierarchical Clustering}
\label{sec:discrepancy-dendrogram:clustering}

Flat partitioning cluster algorithms, like $k$-means, divide the dataset into an a-priori specified number of groups while minimizing intra-group distances. On the other hand, hierarchical clustering algorithms retain all cluster structures in the dataset and, therefore, do not require a $k$ parameter. Hierarchical clustering is thus preferable because it will contain all possible clusters the analyst might be interested in, and we can enumerate them. We chose a clustering by agglomerative nesting (AGNES) \cite{rohlf1982} because bottom-up hierarchical clustering is easier to think about and, thus, easier to explain to analysts than the top-down variant. Further, many current alternatives, such as HDBSCAN \cite{campello2013}, require Euclidean distances and can not be used with just dissimilarities. The main parameter of AGNES is the linkage criterion, i.e., how to compute the distance between two clusters. Only some linkage criteria \emph{can} be used in our case. E.g., centroid-based variants like Ward's method are not applicable as the concept of a centroid may not exist for complex data types, such as regionalizations. Consequently, we provide complete and average linkage as user-selectable hierarchical clustering parameters.

\subsection{Normalize Cluster Distances}
\label{sec:discrepancy-dendrogram:distance-normalization}

We aim to evaluate whether a given cluster shrinks or expands when an alternative dissimilarity measure $d_A()$ is applied. The obvious problem here is that $d()$ and $d_A()$ might have differing images, i.e., one maps to the unit interval $[0,1]$ while the other maps onto $[0,1312]$. We propose ranking or min-max normalization to solve this issue. Both operations work on a distance matrix. Ranking replaces values in all cells by their rank, while min-max normalization maps values onto the unit interval. When comparing ranks, the focus will naturally be on ordinal changes, ignoring magnitude. Min-max normalization, on the other hand, preserves magnitude. The analyst can switch between the two as both approaches have advantages and drawbacks (compare \Fref{fig:evaluation:sbss:co-kw}).

\subsection{Compare Cluster Diameters (Sensitivity Index)}
\label{sec:discrepancy-dendrogram:cluster-comparison}

Finally, we require a way to measure a cluster's diameter, which roughly corresponds to the linkage criterion in \Fref{sec:discrepancy-dendrogram:clustering}. To find candidates, we turn to internal clustering validation measures \cite{liu2010}, as no external information exists in our case. These usually incorporate the \emph{compactness} of clusters, which measures the variation within a cluster. Based on the selected linkage criterion, we use the largest distance between any two elements (complete linkage) or the average distance between all elements (average linkage).

Given two distance-normalized hierarchical clusterings P and O (e.g., one with distances of parameter settings and one with output distances) and a cluster diameter definition, we can compute by how much a cluster in P shrinks or expands in O, or the other way around, as P and O cluster the same data cases. We evaluate the \emph{index()} function (Alg.~\ref{alg:si}) for every cluster, i.e., every horizontal line in a dendrogram. $D_{\{1,2\}}$ are the respective distance matrices of P and O. The subroutine \texttt{upperTri} returns the upper triangle of a square matrix, and \texttt{select} selects specified rows and columns of a square matrix. The function can be seen as a sensitivity index as it quantifies how much the variation differs between the parameter and output space.


\begin{algorithm}[tb]
\DontPrintSemicolon
\caption{Pseudocode of sensitivity index computations.}
\label{alg:si}
\KwData{Cluster $C$ of data cases, normalized distance matrices $D_1$ and $D_2$, cluster diameter definition $diam()$.}

\SetKwFunction{upperTri}{upperTri}
\SetKwFunction{select}{select}
\SetKwFunction{diam}{diam}

\SetKwProg{Fn}{Function}{ is}{end}

\Fn{index($C$, $D_1$, $D_2$, $diam$)}{
$D_1[C] \gets \upperTri(\select(D_1, C))$\;
$D_2[C] \gets \upperTri(\select(D_2, C))$\;
\Return{$diam(D_1[C]) - diam(D_2[C])$}

}
\end{algorithm}

\subsection{Visualization}
\label{sec:discrepancy-dendrogram:visualization}

Two established visualization idioms for clusterings are dimensionally-reduced scatterplots and dendrograms. As our target users (statistics experts) found the former approach in previous work \cite{piccolotto2022a} rather unintuitive, we chose the latter for our context, fulfilling design guideline D3 (\Fref{sec:task-abstraction}). Additionally, a dendrogram supports many other guidelines and user tasks. The leaves are juxtaposed (D2), and similar leaves, which are joined into clusters earlier than dissimilar leaves, naturally appear adjacent (D1).
Optimal leaf orderings may be used \cite{bar-joseph2001}.
Lines encode the diameter of every possible cluster that could be interesting (T2--T3). These lines do not overlap (D2). The open challenges are encoding the sensitivity index (\Fref{sec:discrepancy-dendrogram:cluster-comparison}) in the dendrogram (T4) and ensuring that visualizations of data cases are visible (T5).

The free visual channels in a dendrogram we could use to support T4 are line color (hue, saturation), line texture (e.g., dashed or dotted), and line thickness. We encoded the sensitivity index in color hue (compare \Fref{fig:teaser}). The index diverges with 0 at the center. Hence, the direction is as important as the magnitude. Two-directional encodings are standard for color hue (diverging scales) but very uncommon for the other attributes and likely confusing for our target users. We use two diverging scales dependent on the choice of distance normalization (\Fref{sec:discrepancy-dendrogram:distance-normalization}): Red--blue (ranked) and purple--green (min-max). By default, the color scale spans the whole theoretically possible index interval, but the analyst may instead use the interval as found in the dataset to highlight small-scale patterns.

To support task T5, we show customized space-efficient visualizations as leaves of the \dd (\Fref{fig:prototype}-A, bottom). There is little available space when the dendrogram shows many data cases. We combat this issue with several strategies. First, clusters of the dendrogram can be hidden. Second, when leaves are clicked, a tooltip containing a more detailed visualization appears. I.e., we show the regionalization parameter of SBSS as flat polygons in the dendrogram and as an interactive Leaflet map in tooltips. Any cluster can be selected to be shown in the Gallery (\Fref{fig:prototype}-B). More interactions are described in \Fref{sec:visualizations:dendrogram}.

\subsection{Interpretation, Notation and Example}

\begin{figure}[tb]
    \centering
    \includegraphics[width=.8\linewidth]{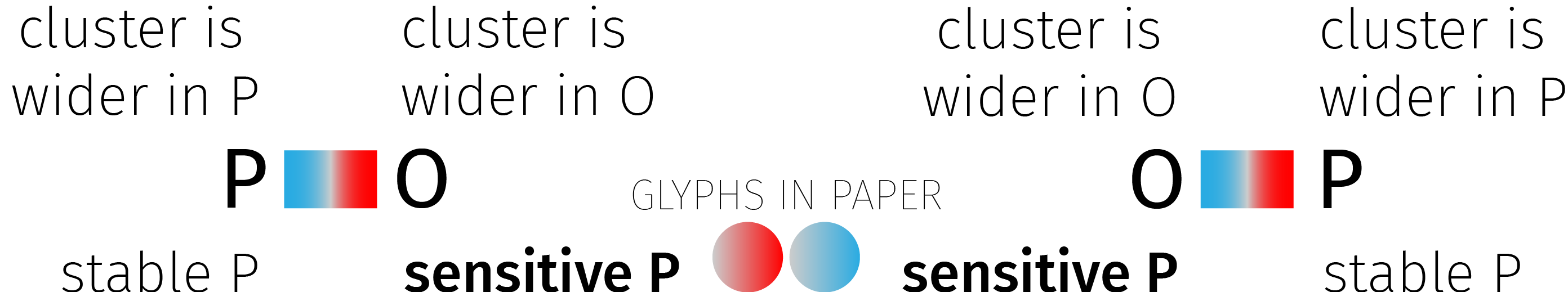}
    \caption{Parameter assessment changes depending on the assignment to primary and alternative distance in the \dde. Glyphs in the document show the color of wider output clusters.}
    \label{fig:glyphs}
\end{figure}

The choice of the color scale's orientation is arbitrary. We decided that red (purple) highlights an expanded cluster while blue (green) marks shrunk clusters in the alternative distance (O in \Fref{fig:teaser}). Consequently, interpretations regarding stability or sensitivity depend on how parameters and outputs are assigned to primary and alternative distances (\Fref{fig:glyphs}). E.g., sensitive parameter settings are associated with wider clusters in the output space compared to the parameter space, which can appear as blue (parameter as primary distance) or red (parameter as alternative distance). In the remainder of the paper, we will use appropriate glyphs to denote the direction of sensitive parameters. A \glyphSenRede\ddshorte{X}{Y} \dd will thus i) compare X and Y, ii) show a dendrogram of clusters in X, iii) mark data cases with sensitive parameter settings as red.

\begin{figure}[tb]
    \centering
    \begin{tikzpicture}
        \node[inner sep=0pt] (app) at (0,0) {\includegraphics[width=\linewidth]{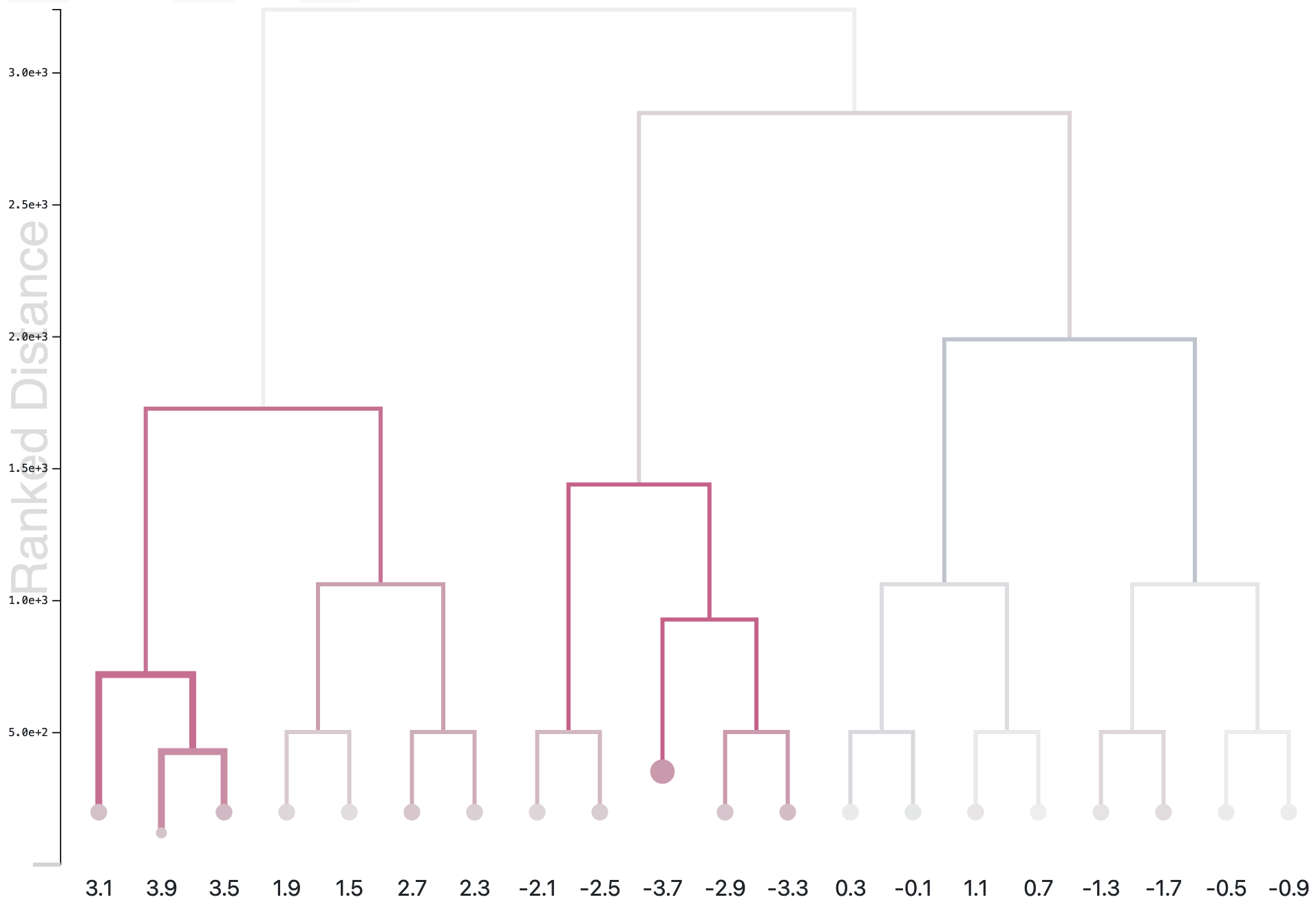}};

        \node[anchor=south west, fill=white] (bg) at (.65,.25) [draw,thin,minimum width=3.5cm,minimum height=2.6cm] {};

        \node[inner sep=0pt] (graph) at (0.7,0.6) {
            \begin{axis}[
                axis lines = left,
                axis background/.style={fill=white}
            ]
            \addplot [
                domain=-4:4,
                samples=16,
                color=black,
            ]
            {x^2};

            \end{axis}
        };
        \node[anchor=south west, fill=red, very nearly transparent] (left) at (1.2, .62) [draw,thin,minimum width=.75cm,minimum height=2cm] {};
        \node[anchor=south west, fill=red, very nearly transparent] (right) at (3.1, .62) [draw,thin,minimum width=.75cm,minimum height=2cm] {};
        \draw[thick,gray,-to] (-2.25,-.2) -- (3.3,1.8);
        \draw[thick,gray,-to] (-.25,-.5) -- (1.5,1);
        \draw[thick,gray,-to] (2.75,-1.5) -- (2.75,1.25);
    \end{tikzpicture}

    \caption{\glyphSenRede\ddshorte{X}{Y} \dd of the function $y=x^2$ (inset top right), with some clusters collapsed for readability. Red color highlights clusters that are wider in $Y$ than $X$ (=sensitive parameter ranges, i.e., marked parabola arms in inset).}
    \label{fig:app:example}
\end{figure}

\Fref{fig:app:example} shows a \glyphSenRede\ddshorte{X}{Y} \dd for the function $y=x^2$ sampled uniformly in the interval $[-4,4]$. The dendrogram separates the parameter space into three clusters with $X$ $1.4$ to $4$, $-4$ to $-1.9$, and $-1.8$ to $1.3$ (from left to right). The lines' hue may be interpreted as the absolute gradient: Red lines mark wider clusters in Y (high) while the right-most cluster is gray (low). When plotted as a line chart, these patterns would refer to the parabola arms (red clusters) and the part between them (gray) as visible in the inset.

\section{Visual Analytics Prototype}
\label{sec:visualizations}

\begin{figure*}
    \begin{tikzpicture}
	\node[anchor=south west,inner sep=0] (image) at (0,0)  {\includegraphics[width=\linewidth]{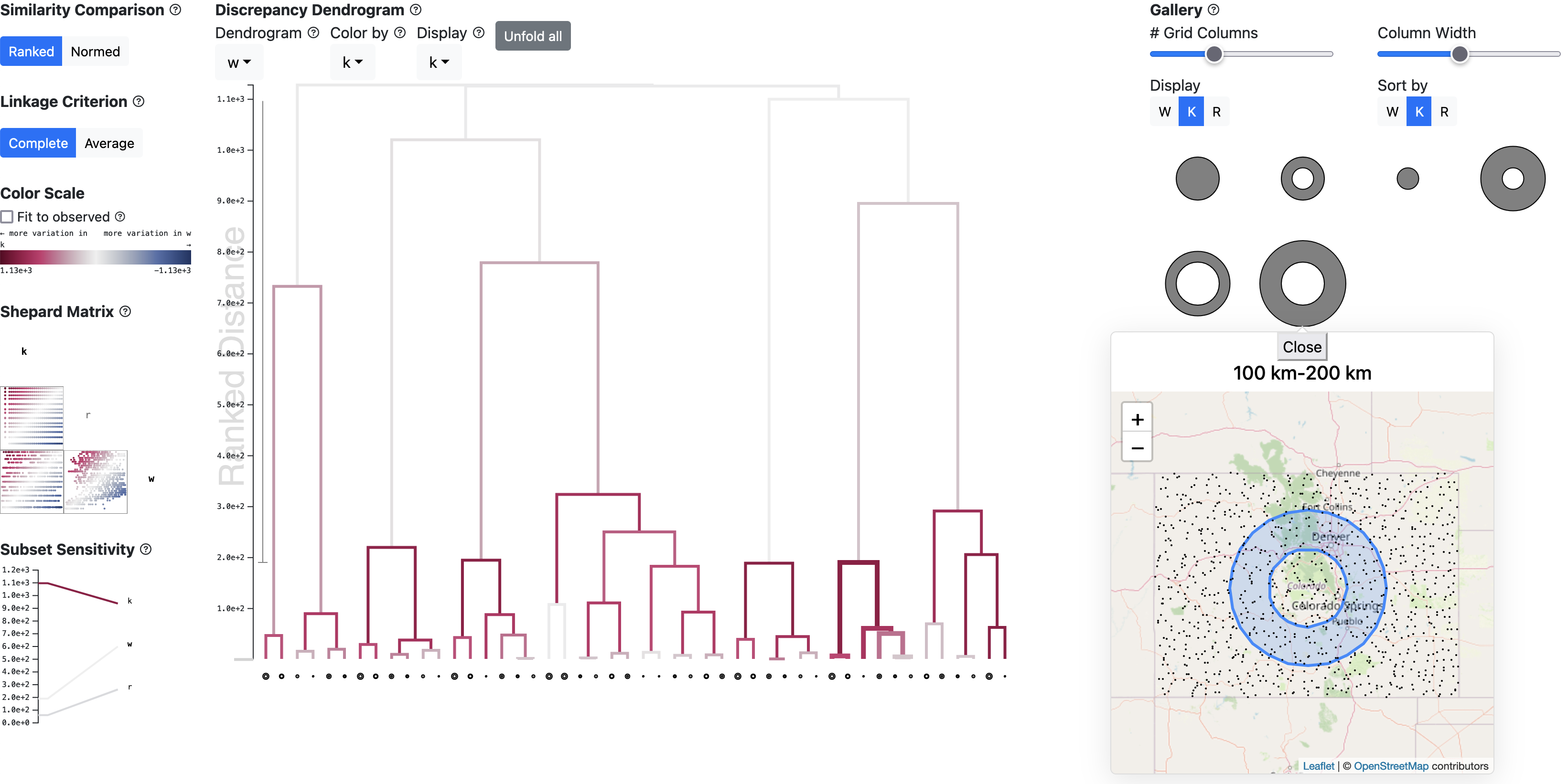}};
    \begin{scope}[x={(image.north west)},y={(image.south east)}]
		\node [textWcolor, anchor=north west] (a) at (0.750, 0.4) {A} ;
		\node [textWcolor, anchor=north west] (b) at (0.975, 0.7) {B} ;
		\node [textWcolor, anchor=north west] (c) at (0.300, 0.1) {C} ;
		\node [textWcolor, anchor=north west] (d) at (0.600, 0.1) {D} ;
		\node [textWcolor, anchor=north west] (e) at (0.550, 0.7) {E} ;

		\node [textWcolor, anchor=north west] (a1) at (0.875, 0.150) {A1} ;
		\node [textWcolor, anchor=north west] (a2) at (0.875, 0.220) {A2} ;
		\node [textWcolor, anchor=north west] (a3) at (0.875, 0.270) {A3} ;
    \end{scope}
    \end{tikzpicture}

  \centering
  \caption{Screenshot of our prototype showing 48~SBSS parameters and outputs (\Fref{sec:evaluation:sbss}). Components: (A) \dd (\Fref{sec:discrepancy-dendrogram}, \Fref{sec:visualizations:dendrogram}), (B) Gallery (\Fref{sec:visualizations:gallery}), (C) Subset Sensitivity View (\Fref{sec:visualizations:subset-sensitivity}), (D) Shepard Matrix (\Fref{sec:visualizations:splom}), (E) tooltip.}
  \label{fig:prototype}
\end{figure*}

To facilitate \sa of SBSS parameters and outputs, we propose a visual analytics prototype (\Fref{fig:prototype}). We developed it in a user-centered design process in collaboration with statistics experts, who are co-authors of this paper. Links to a web version of the software are available in the supplemental material.

\subsection{\dd (T2--T5, D1--D3)}
\label{sec:visualizations:dendrogram}

We discuss the construction of the \dd in \Fref{sec:discrepancy-dendrogram} and focus here on interactions. We provide several interactions with the \dd to allow detailed investigation of clusters and to scale it to larger datasets.
First of all, the user may choose between ranked and normalized distances (\Fref{sec:discrepancy-dendrogram:distance-normalization}) and select the bounds of the color scale (\Fref{fig:prototype}, left top). Further, they may choose the linkage criterion for the dendrogram (\Fref{sec:discrepancy-dendrogram:clustering}), which also affects cluster diameter computations (\Fref{sec:discrepancy-dendrogram:cluster-comparison}).
Second, there might be multiple parameters and outputs in a given dataset. The analyst can thus select which parameter/output to build the dendrogram with (primary distance), which parameter/output to compare it to in the sensitivity index calculation (alternative distance), and which parameter/output to show in the dendrogram leaves (\Fref{fig:prototype}-A1 to A3). As color hue is not a precise visual channel, we encode a cluster's diameter difference additionally in the length of a vertical line segment next to the dendrogram's Y-axis legend.
Other dendrogram interactions are more concerned with scalability. It is possible to collapse a cluster (shift + click), collapse all other clusters (meta + click), and collapse all clusters below a user-defined height (click on the Y-axis). These interactions free up additional display space for an area of interest. Colored circles replace collapsed clusters. The circle size is proportional to the amount of data cases in the cluster, while the color corresponds to the clicked line's color. The data cases of a collapsed cluster are replaced by a cluster representative.
Finally, a cluster can be selected, after which contained data cases are shown in the Gallery (\Fref{fig:prototype}-B).

\subsection{Gallery (T5)}
\label{sec:visualizations:gallery}

The Gallery shows data cases of a selected cluster in a grid (\Fref{fig:prototype}-B). The number of columns and their width can be selected by the analyst, as can the sort order of data cases and which parameter or output they should show. It is possible, e.g., to sort parameter visualizations by output similarity, as is often done in visual parameter space analysis \cite{luboschik2014, eichner2020}. Thus, the Gallery can show complex patterns.

\subsection{Subset Sensitivity View (T4)}
\label{sec:visualizations:subset-sensitivity}

The Gallery shows data cases of a selected cluster in a grid (\Fref{fig:prototype}-B). The number of columns and their width can be selected by the analyst, as can the sort order of data cases and which parameter or output they should show. It is possible, e.g., to sort parameter visualizations by output similarity, as is often done in visual parameter space analysis \cite{luboschik2014, eichner2020}. We obtain the sort order by a 1D multidimensional scaling projection. Thus, the Gallery can show complex patterns.

\subsection{Shepard Matrix (T1)}
\label{sec:visualizations:splom}

We want to give analysts a way to judge which parameter-output relations to investigate (T1). To this end, we use a Shepard diagram \cite{deleeuw2005} showing all pairwise distances of data cases in a scatterplot. Each axis is the distance according to one measure. A diagonal line in a Shepard diagram thus means a perfect correspondence between two distance measures, and a dispersed Shepard scatterplot may be more interesting to investigate.
We use the same color hue as in the \dd for dots in a Shepard diagram, i.e., the further away from the diagonal, the more color hue is used. As the dataset usually has more than two parameters/outputs, we adapt the scatterplot matrix to Shepard diagrams to show all possible combinations (\Fref{fig:prototype}-D).

\section{Evaluation}
\label{sec:evaluation}

We evaluated our visualizations heuristically and with expert interviews. The TU Wien pilot ethics board assessed our methods. Thus, our research adheres to the highest ethical standards. Specifically, our research questions were:

\begin{itemize}
    \item (RQ1) Does our visualization design allow efficient and effective \sa for SBSS parameters/outputs?
    \item (RQ2) Is our designed guidance effective?
    \item (RQ3) Does our visualization design transfer to other contexts than SBSS?
\end{itemize}

For RQ1 and RQ2, we conducted a heuristic evaluation with five visualization experts (\Fref{sec:evaluation:vis}). Two SBSS experts used our visualizations on their own data (\Fref{sec:evaluation:sbss}), which also informs RQ1 and RQ2. Finally, for RQ3, we discussed visualizations with a microclimate simulation expert using an appropriate dataset. In this section we use two-letter shortcuts for people: Just letters indicate authors (e.g., \npe) and a trailing number refers to participants (e.g., \mve).

\paragraph*{Procedure.} All sessions started with a 30~minutes introduction where we explained our problem context and the visualizations independently from the available datasets in the prototype. The slides are available in the supplemental material. After the introduction, visualization experts continued with the questionnaire. The other experts used the prototype on a dataset and parameter settings they were familiar with. A semi-structured interview followed for all participants.

\subsection{Visualization Experts (RQ1, RQ2)}
\label{sec:evaluation:vis}

We evaluated our visualization design heuristically with visualization experts according to the ICE-T method \cite{wall2019}. While a good design does not imply that the visualizations are effective, we think the inverse most likely holds (bad design $\rightarrow$ ineffective). Our chosen method is a good compromise between insights gained and the time requested from participants. We asked five participants (four Ph.D. students and one post-doc) from various universities to join our evaluation. We mostly met them over Zoom, and the sessions took around one hour each. According to ICE-T guidelines, five people are sufficient.
Participants were free to use the prototype with various datasets on their own computers. They could always return to the visualization while filling out the ICE-T questionnaire. ICE-T responses are on a 7-point Likert scale. We asked them to share their thought process to understand their critique better. 

\begin{table}[tb]
    \centering
    \begin{tabular}{l  | r p{2.5cm} | r}
        \multicolumn{1}{c}{\textbf{Component}} & \multicolumn{2}{c}{\textbf{Mean}}  & \multicolumn{1}{c}{\textbf{Std.dev}} \\ \midrule
        Insight & 6.26 & {\small \textsf 1} \begin{tikzpicture}[]
            \draw[gray, thin] (-1,0) -- (1,0);
            \fill[] ($(-1,0) + (2/7*6.26,0)$) circle (.05);
        \end{tikzpicture} {\small \textsf 7} & 1.06 \\
        Confidence & 5.11 & {\small \textsf 1} \begin{tikzpicture}[]
            \draw[gray, thin] (-1,0) -- (1,0);
            \fill[] ($(-1,0) + (2/7*5.11,0)$) circle (.05);
        \end{tikzpicture} {\small \textsf 7} & 2.03 \\
        Essence & 5.32 & {\small \textsf 1} \begin{tikzpicture}[]
            \draw[gray, thin] (-1,0) -- (1,0);
            \fill[] ($(-1,0) + (2/7*5.32,0)$) circle (.05);
        \end{tikzpicture} {\small \textsf 7} & 1.45 \\
        Time & 6.08 & {\small \textsf 1} \begin{tikzpicture}[]
            \draw[gray, thin] (-1,0) -- (1,0);
            \fill[] ($(-1,0) + (2/7*6.08,0)$) circle (.05);
        \end{tikzpicture} {\small \textsf 7} & 0.91 \\  \midrule
        \textbf{Total} & \textbf{5.83} & {\small \textsf 1} \begin{tikzpicture}[]
            \draw[gray, thin] (-1,0) -- (1,0);
            \draw[black,thin] ($(-1,-.0625) + (2/7*5,0)$) -- ($(-1,.0625) + (2/7*5,0)$);
            \fill[] ($(-1,0) + (2/7*5.83,0)$) circle (.05);
        \end{tikzpicture} {\small \textsf 7} & \textbf{1.50}
    \end{tabular}
    \caption{Results of the ICE-T evaluation with visualization experts. Responses were on a 7-point Likert scale. A total mean greater than five (small bar) is considered a success.}
    \label{tab:ice-t}
\end{table}

\Fref{tab:ice-t} holds the results of these questionnaires, split by ICE-T component. The complete responses are available as supplemental material. Wall et al.~\cite{wall2019} state that a visualization design is successful when the mean score exceeds five, which we clearly achieved with an overall mean of 5.83.
Our visualization's worst-scoring component (mean 5.11) is Confidence, which is also the one with the highest standard deviation. While participants agreed that we use \enquote{meaningful and accurate visual encodings} (question Q18 in the ICE-T questionnaire) and \enquote{avoid misleading representations} (Q19), they mostly disagreed that our visualization \enquote{promotes understanding beyond individual data cases} (Q20) or highlights data quality issues (Q21). It would take some effort to detect duplicate or invalid data cases in our visualization, but that was a conscious design choice. The second-worst component is Essence, which also has the second-highest standard deviation, indicating disagreement between participants. In fact, the two most contested questions here were whether the visualization \enquote{facilitates generalizations and extrapolations} (Q16) or \enquote{helps understand how variables relate in order to accomplish different analytic tasks} (Q17). Low ratings in the former were, e.g., because the \dd assesses individual clusters but does not indicate differences between elements. This issue could be tackled in the future by specially crafted comparison visualizations. In the latter question, some participants focused on the \enquote{\emph{different} analytic tasks} and argued that our visualization does not fulfill this criterion due to its singular focus.


On the other hand, participants rated the Insight and Time components very well. Two questions of the former seemed somewhat controversial, as they are associated with higher standard deviations (1.79 and 1.64). One participant \emph{somewhat disagreed} that the visualization \enquote{facilitates perceiving relationships in the data} (Q2). Their reasoning was as follows. We show data cases as leaves in the \dd and also in a gallery to the side. However, all data cases are separate visualizations, so it would be akin to showing individual bars instead of a histogram. However, they also realized that this was not a goal of our visualization design. The other contested question was whether the visualization \enquote{helps identify unusual or unexpected, yet valid, data characteristics} (Q5). One participant \emph{somewhat disagreed,} mentioning that data cases with unusual or unexpected features would be hard to spot if the distance metrics would not consider these. We do not see this as an issue because the chosen dissimilarity metrics might as well measure local differences.

\subsection{SBSS (RQ1, RQ2)}
\label{sec:evaluation:sbss}

Two experts (\cc and \jve) in statistics and SBSS, who were not part of the design process, used our visualizations on familiar datasets. They were recruited from the authors' professional network as they were required to have knowledge of SBSS. They both hold a Ph.D. in statistics and published on spatial data analysis. Sessions took around 2~hours. We guided them in the process as much as necessary, e.g., formulated possible analysis goals and answered any questions they had. After that, we continued with a semi-structured interview, inquiring about their confidence in findings, possible insights, and how these relate to prior expectations.

\paragraph*{Datasets and Parameter Settings.} The experts used two spatial datasets. \jv worked on the \emph{Colorado} dataset, which is a geochemical survey of 960~locations and 27~variables in Colorado, USA. Both \jv and \np contributed parameter settings to investigate, as was agreed upon prior to the interview. \jv provided an R script to obtain regionalizations (10~slices along four directions) and kernels (0--200 km radii). \np added regionalizations obtained in a prior study \cite{piccolotto2022}. \cce, on the other hand, worked on the meteorological \emph{Veneto} dataset, which consists of 72~locations and 7~variables in Veneto, Italy. Parameter settings were obtained in a pilot session by \cc and \np together using an existing prototype \cite{piccolotto2022}. We computed outputs for a full factorial of selected regionalizations and kernels for both datasets. In total, 42~settings were available for the \emph{Veneto} and 48 for the \emph{Colorado} dataset. 

\paragraph*{Dissimilarity Measures.} We chose appropriate functions together with our collaborators. For the unmixing matrix W, we use the MD-Index \cite{ilmonen2010}, a specialized comparison tool for unmixing matrices. For two kernels (K), we compute the difference of their so-called Spatial Kernel Matrix \cite{muehlmann2022}. We compare two regionalizations (R) by counting location pairs for which the region assignment is not identical.

\begin{figure}[tb]
     \centering
     \begin{subfigure}[b]{0.25\linewidth}
         \centering
         \includegraphics[width=\textwidth]{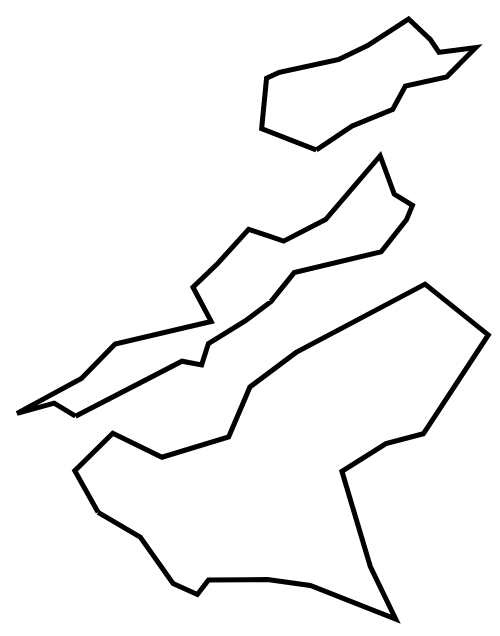}
         \caption{R}
         \label{fig:evaluation:sbss:l-r}
     \end{subfigure}
     \hfill
     \begin{subfigure}[b]{0.25\linewidth}
         \centering
         \includegraphics[width=\textwidth]{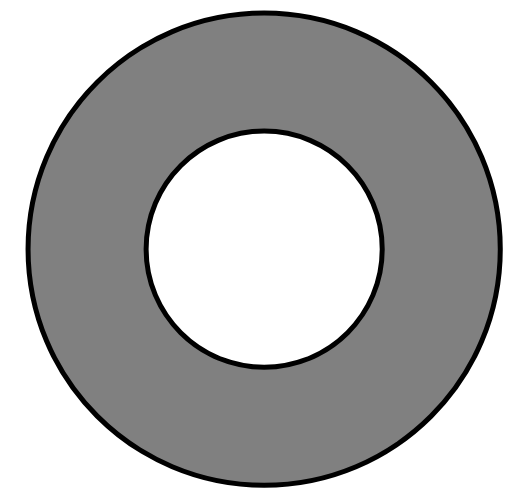}
         \caption{K}
         \label{fig:evaluation:sbss:l-k}
     \end{subfigure}
     \hfill
     \begin{subfigure}[b]{0.35\linewidth}
         \centering
         \includegraphics[width=\textwidth]{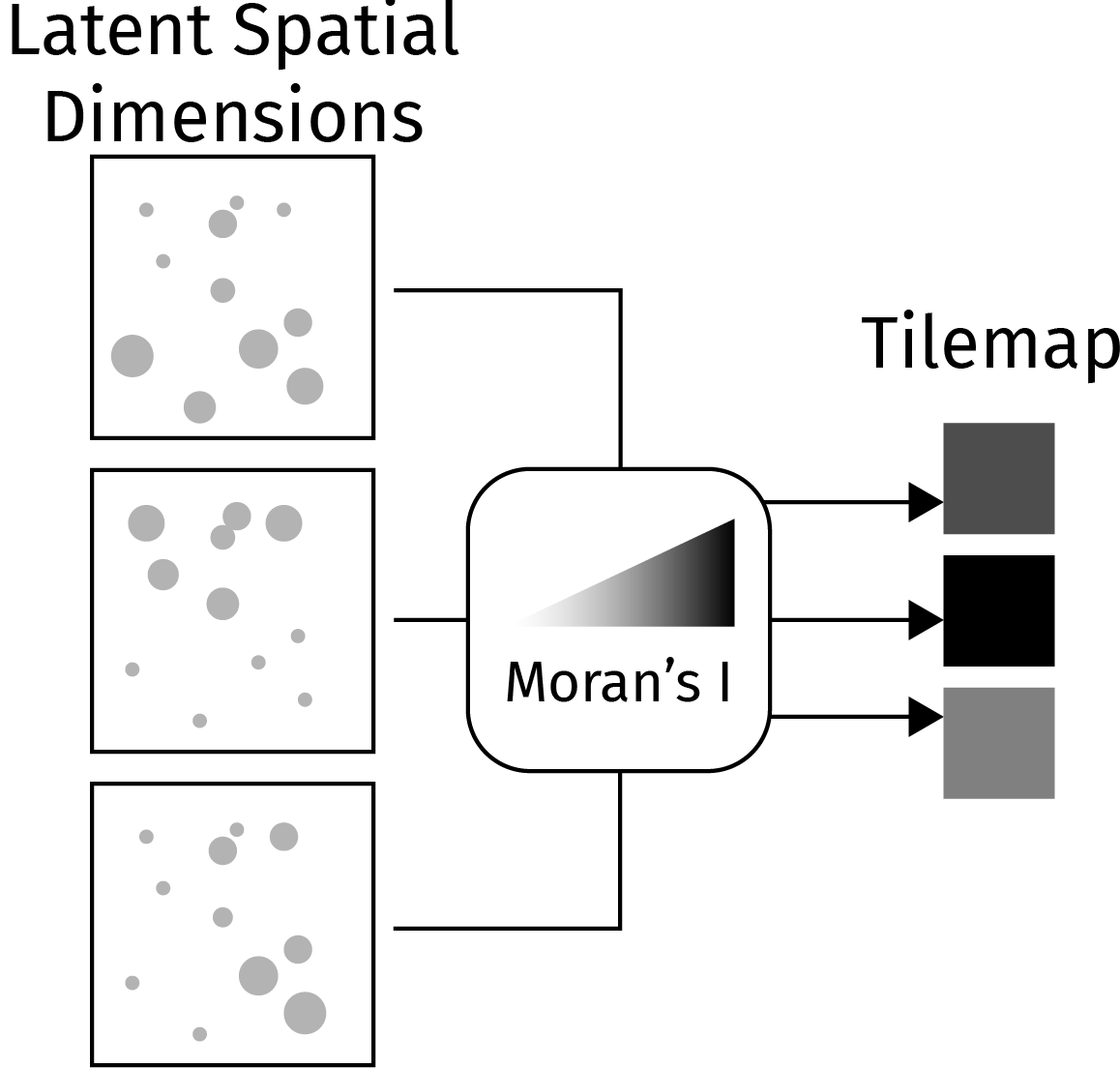}
         \caption{W}
         \label{fig:evaluation:sbss:l-w}
     \end{subfigure}
     \hfill
    \caption{Leaf visualizations for SBSS regionalization (R) and kernel (K) parameter, and output (W).}
    \label{fig:evaluation:sbss}
\end{figure}

\paragraph*{Leaf Visualizations.} We used three visualizations to represent R, K, and W (\Fref{fig:evaluation:sbss}). For R, we showed as multiple polygons representing the concave hull of regions (\Fref{fig:evaluation:sbss:l-r}). In tooltips, these were integrated into interactive Leaflet maps. For K, we showed concentric circles representing the ring size (\Fref{fig:evaluation:sbss:l-k}), also overlaying them to the spatial context with Leaflet in tooltips (\Fref{fig:prototype}-E). We visualized W as a tilemap where each tile represented one latent dimension (\Fref{fig:evaluation:sbss:l-w}). Tiles were colored in a univariate continuous gray color map showing Moran's I \cite{moran1950}, a measure for spatial autocorrelation. High values of that measure point to large-scale spatial patterns, which analysts might find easier to interpret. Tiles were ordered as the SBSS algorithm returned respective dimensions. Tooltips of tiles showed static plots of latent dimensions overlayed on OpenStreetMap.

\begin{figure}[tb]
    \centering
    \begin{tikzpicture}
	\node[anchor=south west,inner sep=0] (image) at (0,0)  {\includegraphics[width=\linewidth]{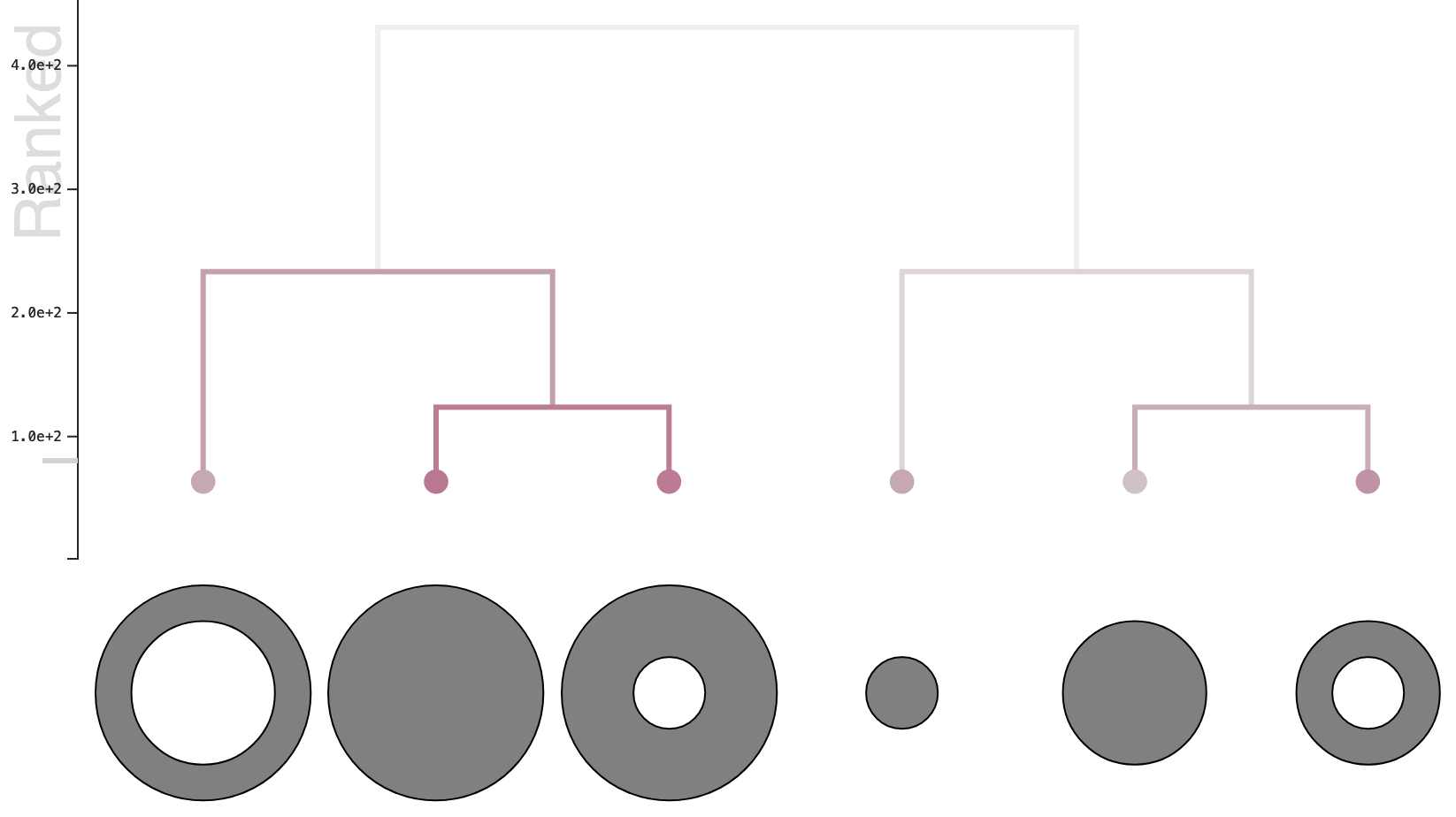}};
    \begin{scope}[x={(image.north west)},y={(image.south east)}]
		\draw[dashed,gray] (.45,.7) rectangle (0,.85);
    \end{scope}
    \end{tikzpicture}
    \caption{\glyphSenRede\ddshorte{K}{W} \dd (average linkage with some clusters collapsed, cropped) for the \emph{Veneto} dataset. The dashed box marks the most stable kernel setting identified by \cce.}
    \label{fig:evaluation:sbss:it-kwk}
\end{figure}

\paragraph*{\cce.} \np guided \cc to focus on \sa because other than \jve, \cc initially focused more on the spatial relationship between regionalizations (R) and locations in the dataset. Regarding \sae, \cc was interested in the influence of the kernel (K) parameter on the output. \np pointed \cc to a \glyphSenRede\ddshorte{K}{W} dendrogram configuration and explained that the red color points to sensitive parameter settings. Almost all K clusters were colored red. As they were wider in W, it indicated that the other parameter (R) exerts more influence on the output than K. \cc switched to average linkage to account for any outliers that may skew the complete linkage criterion. Using this view (\Fref{fig:evaluation:sbss:it-kwk}), they found that K with a radius 0--60~km was the least red compared to others. Hence, this setting was most stable regarding the choice of R, with K=0--30~km a close second. \cc explained that most locations in the dataset are within 75~km, so a kernel up to 60~km will likely capture most of the spatial dependency structure. \cc also observed kernels up to 90~km radius (three big circles on the left in \Fref{fig:evaluation:sbss:it-kwk}) generally showing wider clusters in W than the smaller kernels due to their stronger red color. \cc concluded that two levels of spatial variability exist in the dataset.

\begin{figure}[tb]
    \centering
    \begin{tikzpicture}
	\node[anchor=south west,inner sep=0] (image) at (0,0)  {\includegraphics[width=\linewidth]{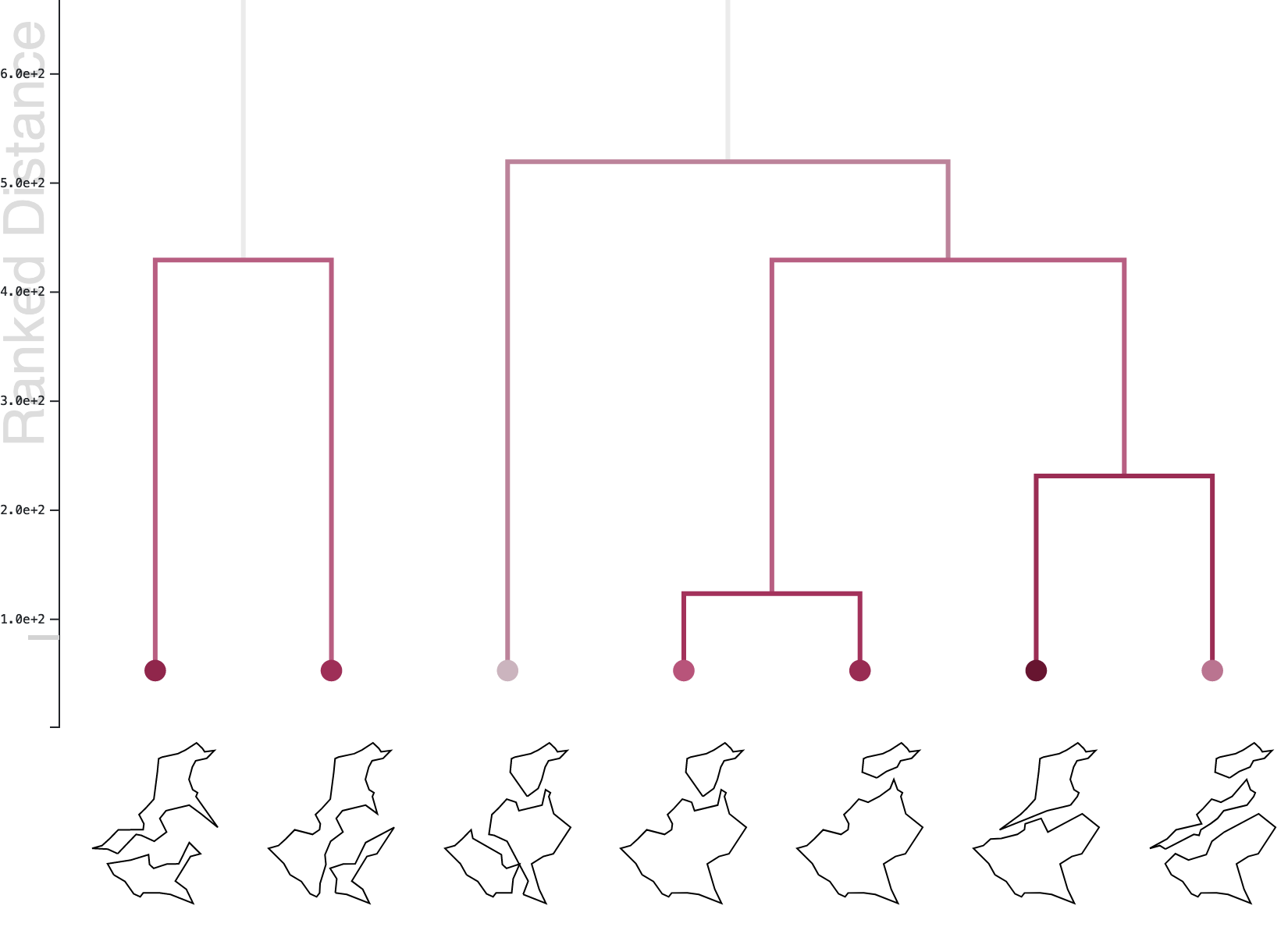}};
    \begin{scope}[x={(image.north west)},y={(image.south east)}]
		\draw[dashed,gray] (.4,.875) rectangle (0,1);
		\draw[dashed,gray] (.4,.325) rectangle (0,.475);
    \end{scope}
    \end{tikzpicture}
    \caption{\glyphSenRede\ddshorte{R}{W} \dd (complete linkage with some clusters collapsed, cropped) for the \emph{Veneto} dataset. Dashed boxes mark the most stable regionalization settings identified by \cce.}
    \label{fig:evaluation:sbss:it-rwr}
\end{figure}

Next, a \glyphSenRede\ddshorte{R}{W} configuration of the \dde, was investigated (\Fref{fig:evaluation:sbss:it-rwr}). Here, clusters of the 3-partitions chosen by altitude and precipitation were the most stable, meaning they were more independent of the choice of K than other partitions. This fact was initially surprising to \cce. However, \cc reconciled it such that the two partitions are similar in that they both separate Veneto's mountainous and flat region. However, another separation in the plane seemed necessary. 2-partitions with just the mountain-flat separation were linked to wider clusters in the output, thus more sensitive to the choice of K.

In the interview, \cc voiced many positive sentiments. They found the visualization \enquote{not difficult} to understand, and the construction of the \dd was logical and easy to follow. \cc liked the interactive maps and that \enquote{you can analyze the data by looking at different aspects in different ways.} \enquote{Half of the work is made [with this tool],} so analysis time is saved compared to the \enquote{classical methods.} In sum, \cc found our visualizations \enquote{help evaluate the parameters} and identified an interesting parameter subspace to consider for future analysis: Smaller K in higher resolutions, as 0--60~km kernels were found to be most stable. \cc could see our visualizations working for people who are \enquote{not completely expert [sic]} in \sae. Based on these sentiments, we think RQ1 and RQ2 can be answered positively.

\cc thought that the \dd is not very easy to interpret but also attributed this to lack of familiarity with our approach and visualizations. Other than \jve, \cc did not confirm or challenge expectations about parameter importance/sensitivity, as they find it necessary to compare multiple datasets before concluding anything. In the same spirit, \cc remarked that a proper data analysis pipeline uses multiple complementing methods, prohibiting sweeping conclusions using our visualizations alone.





\begin{figure}[tb]
    \centering
    \begin{tikzpicture}
    	\node[anchor=south west,inner sep=0] (image) at (0,0)  {\includegraphics[width=\linewidth]{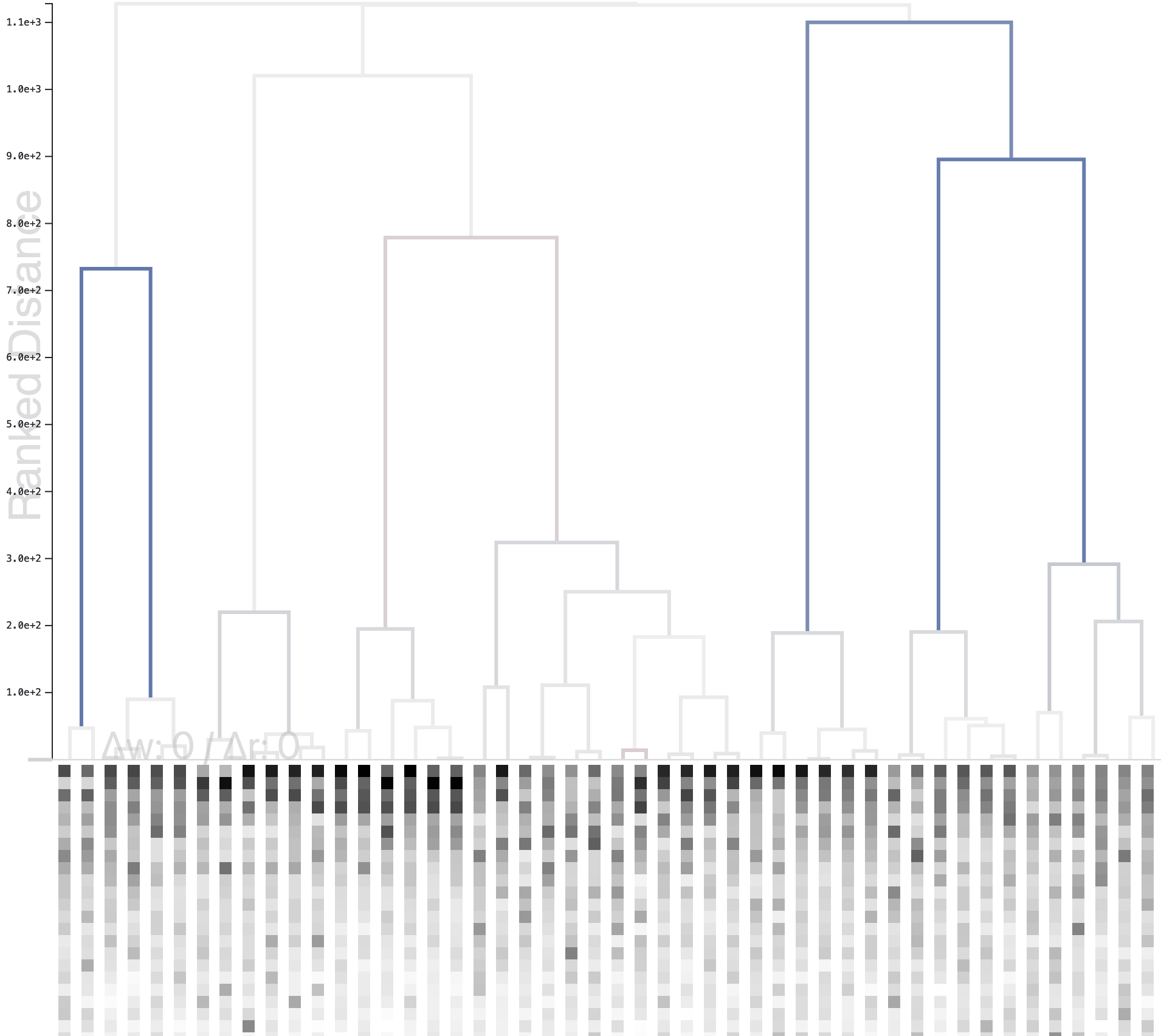}};
        \begin{scope}[x={(image.north west)},y={(image.south east)}]
    		\draw[red,thick,-to] (.8,.2) -- (.75,.15);
        \end{scope}
    \end{tikzpicture}
    \caption{\glyphSenBluee\ddshorte{W}{R} \dd for the \emph{Colorado} dataset. Blue lines mark data cases with variation in W despite similar R. Closer inspection revealed that the presence of a hole of at least 100~km size in associated K settings distinguishes these cases (red arrow).}
    \label{fig:evaluation:sbss:co-wrw}
\end{figure}

\paragraph*{\jve.} First, \jv focused on a \glyphSenBluee\ddshorte{W}{K} \dde. \jv observed many red lines and asked if it was correct to conclude that those outputs are less sensitive to kernel (K) choice, which it was. \jv was then interested in regionalizations (R) and switched to \glyphSenBluee\ddshorte{W}{R}. There, \jv observed a very salient pattern (\Fref{fig:evaluation:sbss:co-wrw}): Most of the dendrogram was gray, indicating that cluster diameters match well between W and R. Thus, R is an important parameter for the \emph{Colorado} dataset. A few clusters showed blue highlights, indicating clusters of sensitive R parameter settings. \jv looked at one of the clusters (red arrow in \Fref{fig:evaluation:sbss:co-wrw}), saw the same R combined with various K, and considered the local dendrogram shape. \jv concluded that two groups of W exist for this R setting (10~horizontal slices): One using very \enquote{un-local} kernels (K) with a 100~km hole and another group containing the dataset's remaining K settings. Hence, the choice of K matters a lot for this particular R setting. Other salient blue patterns were visible on the dendrogram's right side but not investigated by \jve. \jv then returned to the \glyphSenBluee\ddshorte{W}{K} configuration, but set the leaves to show R and investigated how these parameter settings were distributed in the dendrogram. They observed mostly neat clusters (by SBSS output W) of 6~data cases and identical R in each cluster, which was another hint that R is the more important parameter.

\begin{figure}[tb]
    \centering
    \begin{subfigure}[b]{.45\linewidth}
        \centering
        \begin{tikzpicture}
        	\node[anchor=south west,inner sep=0] (image) at (0,0)  {\includegraphics[width=\textwidth]{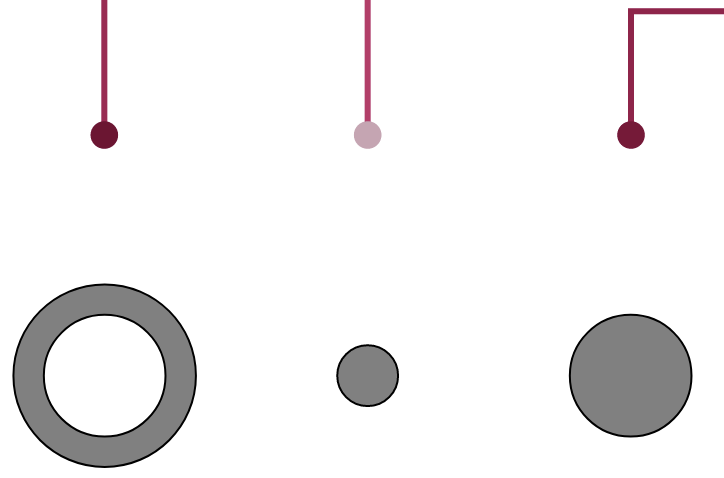}};
            \begin{scope}[x={(image.north west)},y={(image.south east)}]
        		\draw[dashed,gray] (.8,0) rectangle (.6,1);
            \end{scope}
        \end{tikzpicture}
        \caption{Rank}
        \label{fig:evaluation:sbss:co-kw:rank}
    \end{subfigure}\hfill
    \begin{subfigure}[b]{.45\linewidth}
        \centering
        \begin{tikzpicture}
        	\node[anchor=south west,inner sep=0] (image) at (0,0)  {\includegraphics[width=\textwidth]{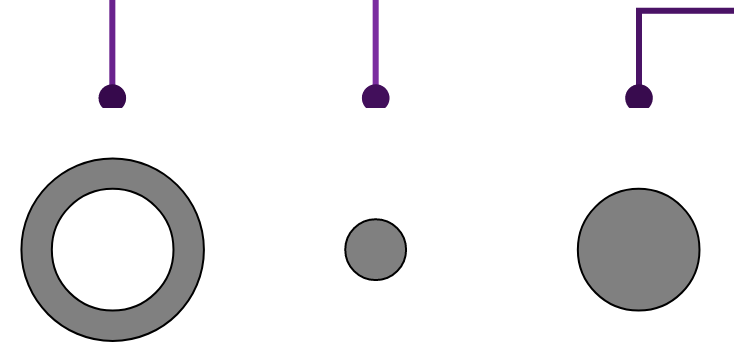}};
            \begin{scope}[x={(image.north west)},y={(image.south east)}]
        		\draw[dashed,gray] (.85,0) rectangle (.6,1);
            \end{scope}
        \end{tikzpicture}
        \caption{Min-max}
        \label{fig:evaluation:sbss:co-kw:minmax}
    \end{subfigure}
    \caption{Rank and min-max distance normalization highlight different relations. Note the enclosed circles' color. With ranked distance (a), the small kernel is shown as most stable of the three (lighter color). Min-max distances (b) show that the absolute difference is low (saturated colors). Our visualizations offer more precise visual encodings in addition to color hue for such comparisons (see, e.g., \Fref{sec:visualizations:dendrogram}).}
    \label{fig:evaluation:sbss:co-kw}
\end{figure}

\np suggested looking at a parameter-focused dendrogram, after which \jv changed it to \glyphSenRede\ddshorte{K}{W}. Here \jv suggested that one K setting (0--50~km radius) is much more stable than the others due to its lighter color and wrongly concluded that R choice matters less for that. While their first assessment (more stable than others) was correct, the second part did not consider the magnitude of the cluster diameter difference in W. If \jv would have used min-max normalized distances \dd (\Fref{fig:evaluation:sbss:co-kw}), they would have seen that also for that K, the cluster diameter difference in W was very high in absolute terms. Finally, \jv also considered \glyphSenRede\ddshorte{R}{W} to investigate the stability of R. Here, a completely different picture than for \glyphSenRede\ddshorte{K}{W} emerged: The lines touching dendrogram leaves were gray instead of red, thus suggesting that less variation happens within R settings than between them. This image was consistent with \glyphSenBluee\ddshorte{W}{R}, underlining the importance of the regionalization (R) parameter even more.

In the interview afterward, \jv offered mainly positive comments. The visualization is \enquote{very intuitive to use}, and it \enquote{speeds up analysis because one can see all parameter combinations at once.} It \enquote{does exactly what it's supposed to do} because \enquote{the color pointed [them] to [data cases] where there was something going on} and, therefore, \jv is \enquote{very confident in results obtained with this tool, I don't doubt it.} They mentioned that \enquote{many observations would not have been possible without this [visualization]} and that, therefore, it can be a qualitative complementary to the quantitative methods they use in their research. E.g., as a more systematic replacement for the trial \& error they do now. \jv confirmed their suspicion that R is the more important parameter using our visualizations. Again, we think these sentiments strongly support both RQ1 and RQ2.

\jv also mentioned that the \dd was not particularly easy to understand. I.e., the syntax, so to say, was clear (red, gray, and blue pointing in the direction of wider clusters), but translating that into actionable steps in parameter analysis was difficult. \jv expects that this effect will get smaller with more familiarity with the visualizations. Finally, \jv admitted they mostly looked at the tilemap in the leaves to judge W similarity. Since tiles contain a summary (Moran's I) of actual maps, the visualization may be misleading. A possible remedy could be a glyph design incorporating a derived feature and map similarity.

\subsection{Microclimate Simulations (RQ3)}
\label{sec:evaluation:vrvis}

To demonstrate that the approach used in our prototype is transferable to other problem contexts (the goal of design studies \cite{sedlmair2012b}), we applied it to microclimate simulation results~\cite{vuckovic2022a}. Such simulation models predict meteorological variables (e.g., air temperature or humidity) in a very small area, typically for a single street or a building. Microclimate simulations are critical nowadays as the climate crisis pressures cities and real estate developers to adapt to changing climate conditions. Usually, stakeholders, like city planners and architects, use existing simulation models and do not develop them themselves. Hence, parameter space analysis so far was mostly done by studying derived features (e.g., maximum temperature) with respect to grid size, often carried out with visual inspection, and computing relations (e.g., correlation) between individual variables. Analysts have certain expectations about parameter relations. These come partially from known model limitations (e.g., the model does not perform well in extreme conditions) and partially from the modeled physical reality (e.g., the humidity of cold vs. hot air or wind chill effects).

In the conducted session, two authors (\npe, \jse) of this paper met with a microclimate simulation expert \mve, who has a Ph.D. in civil engineering and was recruited from the authors' professional network. \js controlled the prototype and suggested findings that \mv assessed, while \np took notes.

\paragraph*{Dataset and Parameter Settings.} The experts' use case was to analyze the climatic conditions around a potential building (available as a 3D model) in several cities, seasons and meteorological conditions (called a \emph{scenario}) to find the best location. The tested cities were Vienna, Helsinki, and Gothenburg in various seasons. \mv computed a dataset containing 12~parameter settings and respective outputs. The low number of data cases follows the simulation model's computational demands as a single run takes several minutes to a couple of hours. The four outputs were wind speed ($O_W$), temperature on the surface ($O_S$) and in the air ($O_A$), and humidity ($O_Q$) at 6~am after a simulated interval of 24~h. The output values are spatially distributed on a grid. Parameters of the model were air temperature ($P_A$) and humidity ($P_Q$) as time series over 24~h, and wind speed and direction ($P_W$). We agreed to use Euclidean distance to measure similarity.



\paragraph*{Leaf Visualizations.} Three visualizations were used to show the model's parameters and outputs, both as leaf and tooltip visualizations. For the spatially distributed outputs ($O_W, O_S, O_A, O_Q$), we used heatmaps with univariate color scales of varying hue. Time series ($P_A, P_Q$) were shown as line charts. Wind speed and direction were shown as arrows, with speed as length and direction as rotation.

\begin{figure}[tb]
    \centering
    \begin{tikzpicture}
    	\node[anchor=south west,inner sep=0] (image) at (0,0)  {\includegraphics[width=\linewidth]{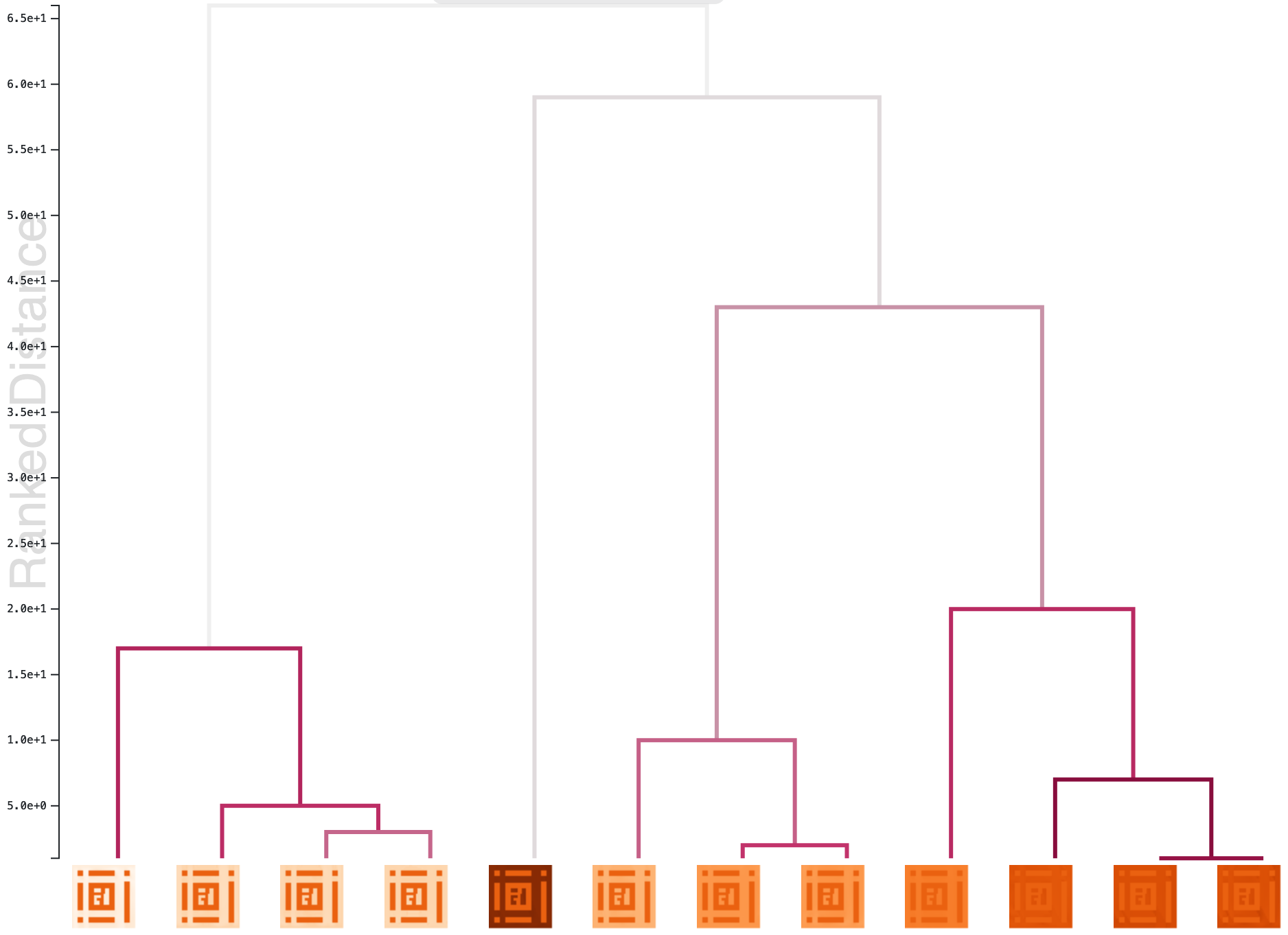}};
        \begin{scope}[x={(image.north west)},y={(image.south east)}]
    		\draw[red,thick,-to] (.55,.3) -- (.45,.4) ;
    		\node[textWcolor] at (.25,.65) {A};
    		\draw[dashed,gray] (.3,.45) rectangle (0,.685);
    		\node[textWcolor] at (.35,.3) {B};
    		\draw[gray,dashed] (.4,0.055) rectangle (0,.35);
    		\node[textWcolor] at (.35,.95) {C};
    		\draw[gray,dashed] (.4,0.7) rectangle (0,1);
        \end{scope}
    \end{tikzpicture}
    \caption{\glyphSenBluee\ddshorte{$O_S$}{$P_W$} \dd used for microclimate simulations. Red lines indicate wider clusters in $P_W$ and thus little influence of that parameter on $O_S$. The red arrow marks a data case suspected to be a model failure. Data cases enclosed by A were also investigated with a \glyphSenBluee\ddshorte{$O_S$}{$P_Q$} configuration. Data cases enclosed by A--C were considered for the final location choice.}
    \label{fig:evaluation:vrvis:model-failure}
\end{figure}

\paragraph{\mve.} In the beginning, we asked \mv about the most important output in the dataset, which \mv answered to be the surface temperature ($O_S$). The goal was to identify a scenario where $O_S$ is both low and stable so as to not be a threat to the human circulatory system. At the same time, general parameter-output relations were of interest. To achieve these tasks, \js set up a \dd with $O_S$ as primary distance and cycled through parameters as alternative distance. We started with a \glyphSenBluee\ddshorte{$O_S$}{$P_W$} configuration, i.e., compared surface temperature output to the wind (direction and speed) parameter. The dendrogram showed many red lines, indicating wider clusters in $P_W$ and thus generally no strong association between $P_W$ and $O_S$. \js changed from ranked to min-max distances to see if the pattern persists when the magnitude is considered, which it did. This relation was expected for \mve.
We also observed an $O_S$ outlier with temperatures up to 36~\textdegree C, which seemed unexpected (red arrow in \Fref{fig:evaluation:vrvis:model-failure}). \mv recalled that \enquote{the simulation model in question aims to capture extreme conditions in summer, like overheating, and there is really the question of how it performs in other conditions and different climates.} \mv concluded that the outlier might be a failure case of the model. Later analysis showed that the presumed model failure was related to extreme temperatures in the $P_A$ parameter. However, it became clear that wind alone \enquote{does not really make a difference} when it comes to surface temperature.

\begin{figure}[tb]
    \centering
    \begin{subfigure}[b]{.45\linewidth}
        \begin{tikzpicture}
        	\node[anchor=south west,inner sep=0] (image) at (0,0)  {\includegraphics[width=\textwidth]{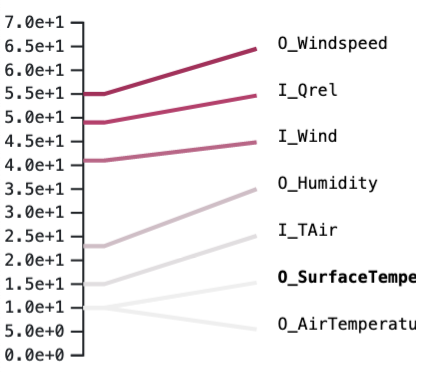}};
            \begin{scope}[x={(image.north west)},y={(image.south east)}]
        		\draw[red,thick,-to] (.39,1) -- (.39,.8);
            \end{scope}
        \end{tikzpicture}
        \caption{Subset Sensitivity View (\Fref{sec:visualizations:subset-sensitivity}) of data cases in \Fref{fig:evaluation:vrvis:model-failure}-A.}
        \label{fig:evaluation:vrvis:subset-sensitivity}
    \end{subfigure}
    \hfill
    \begin{subfigure}[b]{.45\linewidth}
        \begin{tikzpicture}
        	\node[anchor=south west,inner sep=0] (image) at (0,0)  {\includegraphics[width=\textwidth]{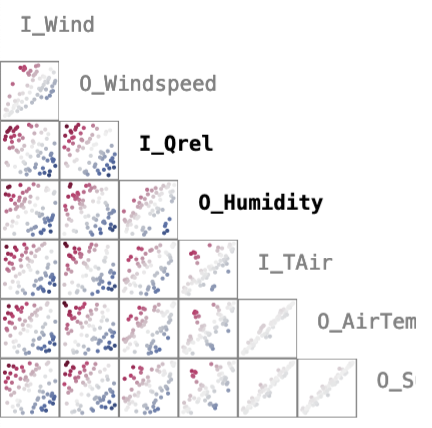}};
            \begin{scope}[x={(image.north west)},y={(image.south east)}]
        		\draw[gray,dashed] (.45,.55) rectangle (0,1);
            \end{scope}
        \end{tikzpicture}
        \caption{Shepard Matrix (\Fref{sec:visualizations:splom}) of microclimate dataset.}
        \label{fig:evaluation:vrvis:splom}
    \end{subfigure}
    \caption{Subset Sensitivity View (a) of cluster in \Fref{fig:evaluation:vrvis:model-failure} shows that air temperature $P_A$ is be the driving parameter for surface temperature $O_S$, as expected. Shepard diagrams of air-related parameters/outputs in Shepard Matrix (b) show that this relation holds for all data cases.}
    \label{fig:evaluation:vrvis:temperature}
\end{figure}

\js then switched to other parameters. Air temperature ($P_A$) was strongly correlated, as expected (\Fref{fig:evaluation:vrvis:temperature}). A similar picture emerged for humidity ($P_Q$), except for a group of three scenarios (\Fref{fig:evaluation:vrvis:model-failure}-A) that arrived at similar $O_S$ with significantly varying $P_Q$ settings. \mv noted that to determine the actual impact of $P_Q$ here, one has to account for the different seasons and cities. This observation was noted as something to investigate later, as, at the time, season and city were not displayed in the prototype. \js then proceeded to compare other outputs with parameters. Our visualizations showed, and \mv confirmed, the known relationship between humidity and air temperature. The next interesting observation came from the connection between wind and temperature. Wind parameter ($P_W$) and output ($O_W$) were not strongly correlated, and air temperature ($P_A$) was identified as another relevant factor (red arrow in \Fref{fig:evaluation:vrvis:subset-sensitivity}). Regarding how temperature could influence wind, \mv mentioned horizontal and vertical mixing effects but that those would be smaller than the wind-to-temperature effects. \mv speculated that some correlations might come from the used 3D grid slices being on pedestrian level (1.8~m) while surface temperature is only valid for the slice at 0~m.


Asked about disadvantages or improvements, \mv mentioned not picking a winning location for their use case because the city and season were missing in the visualization. \np checked together with \js later. Of all the surface temperature ($O_S$) clusters (A--C in \Fref{fig:evaluation:vrvis:model-failure}), $O_S$ was least sensitive to air temperature ($P_A$) in the three scenarios enclosed by A. They belonged to Helsinki (2/3) and Vienna (1). Thus, Helsinki could be identified as the most suitable choice due to the more constant surface temperature. This choice is also consistent with the latest report of the Intergovernmental Panel on Climate Change \cite{gutierrez2021}, which predicts more stable mean temperature for Northern than Central Europe.

To summarize, we could apply our visualizations in a domain they were not originally designed for in the following way. We could find a suitable location for the building, which was the main goal for \mve, thus solving this domain's \sa task. \mv could reconcile visualization images with domain knowledge and find interesting relations to investigate in the future, like the humidity parameter's impact.  We see this session as evidence to support RQ3, that our visualizations can be transferred to other contexts.

\section{Limitations}
\label{sec:limitations}


As we rely on cluster diameters, the particular choice of nested partitions will greatly influence our sensitivity index, visualization image and, ultimately, the analysis outcome. The partitions are in turn influenced by the dataset, dissimilarity measure, clustering algorithm, and its parameters. We took care to select reasonable defaults, but they may not work for every situation. While it may be a demanding task, truthful clusterings can be obtained (cf. \Fref{sec:related-work:clusters}) and the particular groupings could be modifiable by the analyst. Another consequence of relying on relative cluster diameter differences for \sa is that the sensitivity index likely changes when new data is considered, thus the visualization image may be unstable with regard to additions to the underlying dataset. While that may seem like a big constraint, we argue that the same is true for visual \sa of multivariate parameters: If they are sampled too coarsely or in too narrow intervals, then the analysis outcome may change a lot when the previously excluded parameter space is considered.

Our approach (\Fref{sec:discrepancy-dendrogram:cluster-comparison}) roughly corresponds to a one-at-a-time sensitivity index, i.e., a local method. Saltelli et al. \cite{saltelli2019} argue that local methods are only appropriate when the model under investigation is demonstrably linear. We did not confirm whether SBSS (\Fref{sec:evaluation:sbss}) or the microclimate simulations (\Fref{sec:evaluation:vrvis}) are linear models. However, we do not see this as an issue for two reasons. First, local indices in the \sa literature make precise quantitative statements for the whole parameter. As we defined our index only for subsets of data cases, it does not do that. Second, we developed the index for visual guidance in an interactive visualization. As all relevant data cases are visible in detail at any time, the analyst may consider much more context and existing domain knowledge than they would when interpreting only a single number, as demonstrated in \Fref{sec:evaluation}.


\section{Discussion and Conclusion}
\label{sec:discussion}

Based on requirements and observations in the context of SBSS, we developed a data type agnostic approach to visual \sae. It only requires dissimilarity measures and thus works for complex parameters and outputs alike. The core innovation is measuring variation in parameter settings and outputs by cluster diameters. \sa then becomes possible by looking at the difference of the same cluster's diameter in parameter and output space. Evaluation participants expressed high confidence in our visualizations. Future work may improve this paper's proposal by accounting for noise or simultaneously supporting multiple parameters.

The \dd and supporting visualizations (\Fref{sec:visualizations}) were also received very well by evaluation participants, especially considering the task complexity and short training time (around 30~minutes). The construction of the \dd was logical for all participants, and the prototype provided sufficient interactions and levels of detail. The successful heuristic evaluation (\Fref{sec:evaluation:vis}) further supports this evidence. SBSS and microclimate simulation experts could confirm suspected or expected parameter-output relations with our visualizations, while mentioning the need to familiarize themselves more with our approach. E.g., the regionalization parameter R is more important for SBSS than the kernel configuration K (suspected by \jve), or that surface temperature mainly depends on air temperature (expected by \mve). Further, they could make high-level decisions (building location, \mve), find new relevant parameter subspaces (smaller kernels, \cce), or just obtain interesting observations (kernels with holes, \cc and \jve).
Considering the utility of the \dd it will also be interesting to apply our approach to other visualization idioms, e.g., to DR scatterplots (\Fref{sec:related-work:clusters}).

We noted, e.g., during introductory explanations, that some participants found it mentally demanding to reason simultaneously about 1) groups of elements instead of single elements and 2) two distances within a group of elements. This issue is, to some extent, inherent to the problem we want to solve. On the other hand, we think rephrasing \sa or finding visual representations so that analysts can reason about single elements instead of groups has much simplification potential. Achieving this would allow even more powerful \sa visualizations potentially applicable to many contexts (\Fref{sec:evaluation:vrvis}).


\acknowledgments{%
This work was funded by the Austrian Science Fund (FWF) under grant P31881-N32 and the Vienna Science and Technology Fund (WWTF) under grant [10.47379/ICT19047]. We sincerely thank our evaluation participants for their time and expertise.

}

\bibliographystyle{abbrv-doi-hyperref}

\bibliography{template}


\end{document}